%% file: main.tex
\documentclass[sigconf]{acmart}

\def\BibTeX{{\rm B\kern-.05em{\sc i\kern-.025em b}\kern-.08em
    T\kern-.1667em\lower.7ex\hbox{E}\kern-.125emX}}

\renewcommand\footnotetextcopyrightpermission[1]{} 
\setcopyright{none}

\usepackage[utf8]{inputenc}
\usepackage[algo2e,linesnumbered,ruled,vlined]{algorithm2e}
\usepackage{graphicx}
\usepackage{subcaption}
\usepackage{float}
\usepackage{booktabs}
\usepackage{cancel}
\usepackage{amssymb,amsmath,amsthm,amsfonts}
\usepackage{pdfpages}
\definecolor{britishracinggreen}{rgb}{0.0, 0.42, 0.24}

\newcommand{\RNum}[1]{\lowercase\expandafter{\romannumeral #1\relax}}

\newcommand{\NameProblem}[0]{PSD\xspace}

\newcommand{\fl}[1]{\textcolor{purple}{([FL] #1)}}
\newcommand{\sa}[1]{\textcolor{britishracinggreen}{([SA] #1)}}

\newcommand{\hide}[1]{}

\newtheorem{thm}{Theorem}
\newtheorem{lem}{Lemma}
\newtheorem{defn}{Definition}
\newtheorem{observation}{Observation}

\begin{document}
\title[]
{
Towards A Personal Shopper’s Dilemma: Time vs Cost
}
\titlenote{An abridged version of this paper will appear at The 28th ACM SIGSPATIAL Intl Conf. on Advances in Geographic Information Systems 2020 (ACM SIGSPATIAL 2020), Seattle, Washington, USA, November 3-6, 2020.}

\author{Samiul Anwar}
\affiliation{%
  \institution{University of Alberta, Canada}
}
\email{samiul@ualberta.ca}

\author{Francesco Lettich}
\affiliation{%
  \institution{Università Ca' Foscari Venezia, Italy}
}
\email{francesco.lettich@gmail.com}

\author{Mario A. Nascimento}
\affiliation{%
  \institution{University of Alberta, Canada}
}
\email{mario.nascimento@ualberta.ca}

\renewcommand{\shortauthors}{Anwar, Lettich and Nascimento}

\begin{abstract}
Consider a customer who needs to fulfill a shopping list, and also a personal shopper who is willing to buy and resell to customers the goods in their shopping lists.  
It is in the personal shopper's best interest to find (shopping) routes that (i) minimize the time serving a customer, in order to be able to serve more customers, and (ii) minimize the price paid for the goods, in order to maximize his/her potential profit when reselling them. Those are typically competing criteria leading to what we refer to as the \emph{Personal Shopper's Dilemma} query, i.e., to determine where to buy each of the required goods while attempting to optimize both criteria at the same time. 
Given the query's NP-hardness we propose a heuristic approach to determine a subset of the sub-optimal routes under any linear combination of the aforementioned criteria, i.e., the query's approximate linear skyline set.
In order to measure the effectiveness of our approach we also introduce two new metrics, optimality and coverage gaps w.r.t. an optimal, but computationally expensive, baseline solution.
Our experiments, using realistic city-scale datasets,
show that our proposed approach is two orders of magnitude faster than the baseline and yields low values for the optimality and coverage gaps.
\end{abstract}

\maketitle

\input{1_intro}

\input{3_preliminaries}

\input{4_approach}

\input{5_experiments}

\input{2_related}
\input{6_conclusions}

\section*{Acknowledgments}

Research partially supported by NSERC Canada.  F. Lettich started this work while  at the Univ. of Alberta.

\bibliographystyle{unsrt}
\bibliography{biblio}

\appendix
\input{7_appendix}


\end{document}

%% file: 1_intro.tex
\section{Introduction}

\label{sec: intro}

Let us consider a \emph{customer} who needs to acquire some products available in \emph{stores} within a geographical area, e.g., his/her city's road network. 
To accomplish this goal the customer submits his/her request to a \textit{personal shopper} in the form of a \textit{shopping list} of products, along with their respective quantities.

The personal shopper's task is to serve customers by fulfilling their shopping list. 
Serving a customer's request entails two different and \emph{competing} criteria. The first
criterion is the \textit{(shopping) time} needed to serve the customer, i.e., the time needed to visit a sequence of stores in order to acquire all products in the customer's shopping list plus the time needed for delivering them to said customer.  The second one is the \textit{(shopping) cost} of the shopping list, i.e., the sum of the costs of all the products in the shopping list at the stores where they were acquired. 
Clearly, the shopper's main goal should be to minimize \textit{concurrently} both criteria;
the faster he/she can deliver the goods the more customers he/she can serve, and the smaller the actual cost of the goods the higher the profit margins he/she can enjoy.
The main question the shopper needs to answer then becomes:
%
\emph{``Which sequence of stores to visit and which products to acquire in those stores in order to fulfill the customer's shopping list, while, at the same time, minimizing \textit{both} the shopping time and the shopping list's cost?''}
%
Unfortunately, from a practical perspective, it is seldom possible to find a single set of stores that satisfies both criteria. We call this new and practical problem as the \textit{Personal Shopper's Dilemma} (\NameProblem) query. 

At this point one could be tempted to argue that the two considered criteria could be linearly combined into a single one, thus casting \NameProblem into a cost function minimization problem. For instance, though not always desirable, one could express shopping time in terms of monetary cost and then assign appropriate weights to combine both criteria into a single one.  While we use shopping time and shopping cost as the two main criteria one could use, other criteria that could not be expressed in terms of a common unit (e.g., maximize the number of organic and/not locally produced goods, and minimize the customer's waiting time).
There is another, more subtle, limitation to be considered here. Even if a linear combination of the criteria were to be established, appropriate weights would need to be predetermined by the shopper \emph{a priori}. Thus, any -- potentially more interesting -- solution for slightly different combinations of weights would not be returned to the shopper \textit{by design}.  
To overcome such limitations we let 
the shopper decide by him/herself how to (linearly) combine both criteria \emph{a posteriori}.  That is, we aim at computing the set of \textit{all} optimal sequence of stores for \textit{any} linear combination of shopping cost and shopping time, also known as \emph{linear skyline set} (discussed further shortly).
Notably, the shopper's decisions can be made after running \emph{one single} \NameProblem query, instead of running one query for each linear combination of the two criteria.
Thus, in hectic periods of the day the shopper may prefer to prioritize more short trips, each with smaller profit, while calmer parts of the day (or changes in traffic patterns) may push the shopper towards longer trips yielding larger profits, or anything in between.

%
%

\begin{figure}[t]
   \centering
   \includegraphics[width=1\columnwidth]{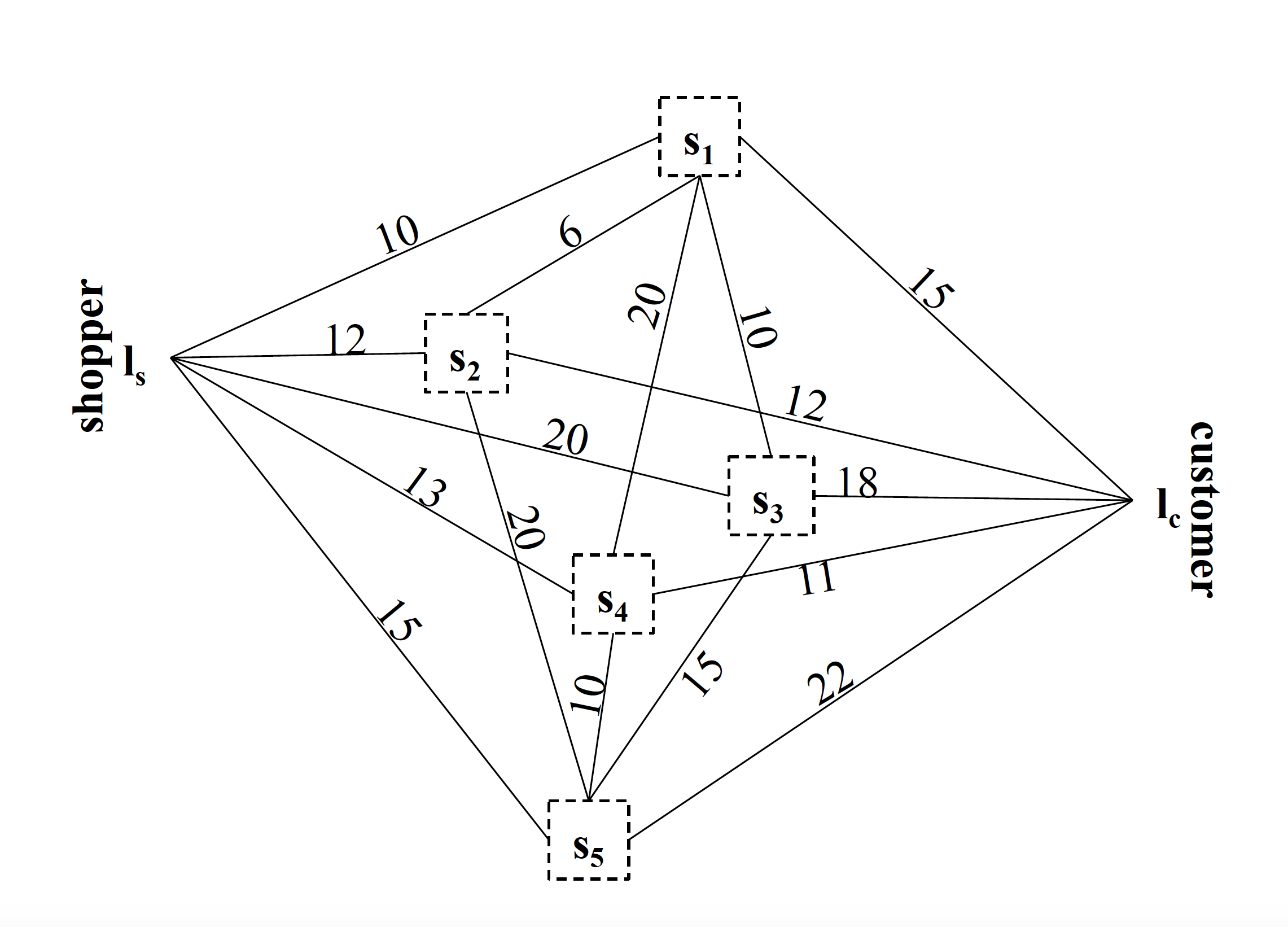}
   \caption{{Sample network.}}
   \label{fig: intro TAHS}
\end{figure}

In order to illustrate the  \NameProblem query, let us introduce a simple instance. Let us suppose that a customer wishes to buy products $A, B, C$ and $D$ and issues the corresponding shopping list to the shopper\footnote{For the sake of simplicity and, in the scope of this example only, let us consider that only one unit of each product is required in the shopping list.}. 
Now, let us suppose that the stores available are those shown in Figure~\ref{fig: intro TAHS} and they are embedded in a road network not shown for the sake of simplicity. Here, stores are depicted by orange squares (and denoted by $s_x$), while the customer's delivery location is denoted by $lc$ and the shopper's current location is denoted by $ls$. Edges between locations represent the fastest paths in the road network connecting them, with labels denoting the associated travel time. Finally, Table~\ref{tab: intro itemlist} shows the products available in each store, along with their unit prices.

\begin{table}[htb]

  \centering
  \begin{tabular}{c l}
    \toprule
    \textbf{Store} & \textbf{Products} \\
    \midrule
    \vspace{2pt}
    $s_1$ & $(A,\: \$7)$, $(B, \:\$8)$, $(F, \:\$10)$\\
    \vspace{3pt}
    $s_2$ & $(C, \:\$10)$, $(D, \:\$8)$, $(E, \:\$10)$\\
    \vspace{3pt}
    $s_3$ & $(C, \:\$5)$, $(D, \:\$4)$, $(F, \:\$6)$\\
    \vspace{3pt}
    $s_4$ & $(C, \:\$8)$, $(D, \:\$7)$, $(F, \:\$12)$\\
    \vspace{3pt}
    $s_5$ & $(A, \:\$6)$, $(B, \:\$7)$, $(E, \:\$8)$\\
    \bottomrule
  \end{tabular}
  \caption{{List of products available in the stores depicted in Figure \ref{fig: intro TAHS}. Each pair represents a specific product and the associated unit cost.}}
 \label{tab: intro itemlist}
\end{table}

From the above scenario we see that there are multiple solutions that may be of interest for the shopper (Table~\ref{tab: intro example}). On the one end of the spectrum route $R_1$ represents the best solution in terms of shopping time, yet its shopping cost is the largest. On the other hand, $R_5$ offers the lowest shopping cost, but it requires to traverse an expensive route in terms of shopping time. Between these extremes there are several solutions that may \textit{interest} the shopper, where the notion of ``interestingness'' depends on the shopper's particular preferences or needs at query time.

\begin{table}[htb]
  \centering
  \begin{tabular}{ccc}
    \toprule
    \textbf{Route} & \textbf{Shopping Time} & \textbf{Shopping Cost} \\
    \midrule
    \vspace{3pt}
    $R_1 = \langle ls, s_1, s_2, lc \rangle$ & 28 & 33\$\\
    \vspace{3pt}
    $R_2 = \langle ls, s_1, s_3, lc \rangle$ & 38 & 23\$\\
    \vspace{3pt}
    $R_3 = \langle ls, s_1, s_4, lc \rangle$ & 41 & 30\$\\
    \vspace{3pt}
    $R_4 = \langle ls, s_5, s_2, lc \rangle$ & 47 & 31\$\\
    \vspace{3pt}
    $R_5 = \langle ls, s_5, s_3, lc \rangle$ & 48 & 21\$\\
    \vspace{3pt}
    $R_6 = \langle ls, s_5, s_4, lc \rangle$ & 36 & 28\$\\
    \bottomrule
  \end{tabular}
  \caption{{Set of routes from Figure~\ref{fig: intro TAHS} that can satisfy the shopping list $\mathbf{\{A, B, C, D\}}$.}}
 \label{tab: intro example}
\end{table}

When dealing with multiple cost criteria and the problem of determining a set of results that are optimal under any arbitrary combination thereof, a well-known and extensively used tool in the literature are \textit{skyline queries} \cite{skyline01}.
In general terms, given a set of cost criteria and a pair of objects $o_i$ and $o_j$, we say that $o_j$ \textit{dominates} $o_i$ if (i) for each cost criterion the cost of $o_j$ is \textit{smaller or equal} than that of $o_i$ and (ii) there is at least one criterion for which the cost of $o_j$ is \textit{strictly smaller} than that of $o_i$.
In turn, the set of objects that are non-dominated by any other object defines the notion of \textit{skyline}, and represents the desired solution. 
If we represent a skyline on a Cartesian plane, then the skyline represents a frontier beyond which all objects are dominated, i.e., they are not better than those in the frontier.

\begin{figure}[t]
   \centering
   \includegraphics[width=0.8\columnwidth]{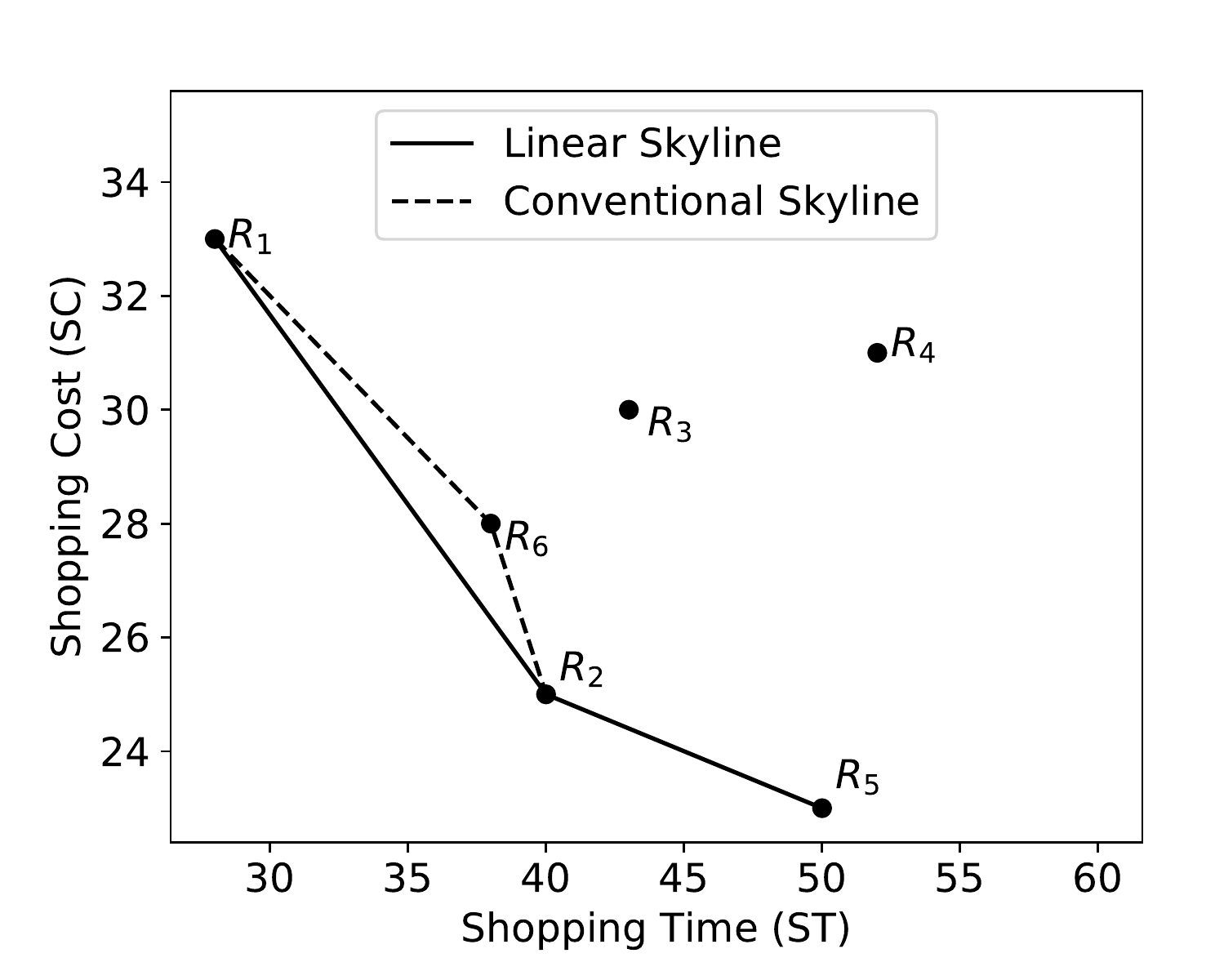}
   \caption{{Comparison between \textit{linear} skyline (continuous curve) vs \textit{conventional} skyline (dashed curve). Shaded area represents the space dominated by the linear skyline.}}
   \label{fig: intro skyline}
\end{figure}

Computing skyline queries is a computationally intensive task that typically returns many very similar solutions, thus possibly making the choice of a specific solution rather difficult for users.
To tackle these issues, Shekelyan et al. \cite{Shekelyan2014} introduced the notion of \textit{linear skylines}. In general terms, a linear skyline is the subset of the objects defining the \textit{convex hull} of the (conventional) skyline. Objects belonging to a linear skyline are required to be optimal under \textit{any linear combination} of the competing cost criteria.  From that, it follows that any solution that is conventionally dominated is also linearly dominated.
Thanks to this stricter, though still practical, requirement, linear skylines typically contain much less objects than conventional skylines and are thus easier to interpret.  
Considering the above example, the linear skyline is represented by the set of routes $LS = \{R_1, R_2, R_5\}$, depicted with a solid line in Figure~\ref{fig: intro TAHS}, whereas the dashed line represents the conventional skyline.

Now we can easily observe the shopper's dilemma.   None of the routes $R_1$, $R_2$ or $R_5$ is strictly better than the others, in that each solution creates a different trade-off between shopping time and shopping cost, and that each can be interesting under different circumstances.
Therefore we target the problem of computing a set of \textit{interesting}, \textit{meaningful}, and  \textit{diverse} shopping routes.

Unfortunately, as formally shown in Section~\ref{app: np-hardness}, even a simplified version of the \NameProblem query, namely one where each store sells a single product and each product is sold at the same price in every store where it is available, is NP-hard. Therefore we propose as our main contribution (in Section~\ref{section:proposed approach}) a heuristic solution that relies on a provably correct \textit{pruning framework} in order to efficiently
retrieve a sub-optimal linear skyline.
We also develop a framework to compare the solutions obtained w.r.t. (costly) optimal ones which we use in our experiments using city-scale realistic datasets (Section~\ref{section:experimental evaluation}).  Overall we found that our proposed heuristic is robust and able to offer solutions of good quality much faster than the optimal approach.

%% file: 3_preliminaries.tex
\section{Preliminaries}
\label{section:preliminaries}

We assume as underlying framework for the \NameProblem query a city's road network modeled by an undirected
graph $G(V,E,W)$, where $V$ is a set of vertices that represent road intersections and end-points, $E$ is the set of edges containing all road segments, and $W$ indicates the weight of edges in $E$. The weight of an edge connecting two vertices $v_i$ and $v_j$, denoted by $w(v_i, v_j) \in W$, is given by the \textit{time} needed to traverse the associated road network segment.

There are four main entities within the \NameProblem's model: stores, customers, shopping lists, and personal shoppers.  
\begin{itemize}
\item 
A \emph{store} $\tau_j$ is located at a vertex $v_{\tau_j} \in V$, and the set of all stores forms a set $T$.
Each store sells a specific selection of products. A product has a positive cost, which may differ between stores, and we denote by 
$c(i,\tau_j)$ the cost of a product $i$ at a store $\tau_j$. For simplicity we assume that all stores have an arbitrarily large inventory of all products it sells.
\item
A single \emph{customer} $\sigma$ wants to buy a set of products. To this end it issues a request to a personal shopper in the form of a shopping list (described next). 
We assume the products need to be delivered at the customer's location, which is a vertex in $V$ denoted by $v_\sigma$.
\item
$\lambda$ denotes the \emph{shopping list} issued by $\sigma$ and we represent it as a set of pairs 
$\langle i, q \rangle$ 
where each $i$ is a product identifier and $q$ represents the number of required units of such product.
\item
Finally, the \emph{personal shopper} $\omega$ is in charge of satisfying a customer's request, i.e., fulfil and deliver the shopping list, and we denote the shopper's location by a vertex $v_\omega$ in $V$. \end{itemize}





The answer of a \NameProblem query is a set of ``shopping routes'', each representing a sequence of stores to be visited and yielding a shopping cost (i.e., the monetary cost of acquiring all required products in the visited stores) as well as shopping time (i.e., the time needed to do all needed shopping and deliver the products to the customer). Next, we define such concepts formally.

\begin{defn}[Shopping route and its feasibility]
A shopping route $\theta_i$ over $G$ is a sequence of stores $\langle \tau_1^i, \ldots, \tau_n^i \rangle$.  
Furthermore, given a shopping list $\lambda$ we say that $\theta_i$ is feasible w.r.t. $\lambda$ if
all products in $\lambda$ are sold in at least one store in $\theta_i$.
\end{defn}


\begin{defn}[Shopping time]
Let $\theta_i$ represent a feasible shopping route w.r.t. a customer's shopping list $\lambda$.
We define the shopping time associated with $\theta_i$, denoted as $ST(\theta_i)$, as the time needed by the shopper to traverse the path that departs from $v_\omega$, visits the stores according to the order defined by $\theta_i$, and finally ends at $v_\sigma$, i.e.:
%
\begin{equation*}
ST(\theta_i) = mTT(v_\omega, \tau_1) + \sum_{i=1}^{n-1}mTT(\tau_i,\tau_{i+1}) + mTT(\tau_n,v_\sigma)
\end{equation*}
where $mTT(v_i, v_j)$ is the time required by the fastest path
connecting vertices $v_i$ and $v_j$ (though other notions of travel cost could also be used).
\end{defn}

\begin{defn}[Shopping cost]
Let $\lambda$ be the shopping list issued by a customer $\sigma$ and 
let $\theta_i$ be a feasible shopping route w.r.t. $\lambda$.
Then, we define the shopping cost of $\theta_i$ as follows:

$$SC(\theta_i) = \sum_{\langle j,q \rangle \in \lambda} (c(j,{ss(\theta_i, j))} \times q)$$
where $ss(\theta_i,j)$ is a function that returns the store in $\theta_i$ from which the product $j$ in $\lambda$ is to be bought~\footnote{As discussed in the following Section, such store-product assignment is determined as the solution is obtained.}.
\end{defn}

Note that due to the assumption of arbitrarily large inventories at each store, it is safe to assume that all units of a given product can be acquired at a single store. The case where stores' inventories are limited, and more than one store may be required to fulfil the need for a given product, is left as future work for the time being.

Clearly, there may be different combinations of stores that could interest the shopper, each with its own trade-off between shopping cost and shopping time.  Thus, we propose an approach where the shopper can do his/her own evaluation to choose the one combination that best fits his/her immediate needs/goals. We model such notion of ``interestingness''  by leveraging the concept of \textit{linear skyline}, discussed next.

A linear skyline always represents the subset of some \textit{conventional} skyline; to distinguish between the two, we start by providing their definitions. Let $\theta_i$ be a feasible shopping route.  The two criteria that are being optimized are $SC(\theta_i)$ and $ST(\theta_i)$, i.e., the cost vector that is being optimized is $CV(\theta_i) = \langle SC(\theta_i), ST(\theta_i) \rangle$.

\begin{defn}[Conventional domination]
\label{def: conventional domination}
Let $\theta_i$ and $\theta_j$ be two shopping routes. Then, we say that $\theta_i$ \textit{conventionally dominates} $\theta_j$, denoting by $\theta_i \prec \theta_j$, if:

\begin{align*}
\Big(\big(SC(\theta_i) < SC(\theta_j) \big) 
\land 
\big(ST(\theta_i) \leq ST(\theta_j)\big)\Big) 
\ 
\lor \\[-0.5em]
\Big(\big(SC(\theta_i) \leq SC(\theta_j)\big) 
\land 
\big(ST(\theta_i) < ST(\theta_j)\big)\Big)
\end{align*}
\end{defn}

From this, it follows the definition of \textit{conventional skyline}.

\begin{defn}[Conventional skyline]
Let $\Theta$ be a set of shopping routes. Then, we define the conventional skyline of $\Theta$ to be the set of shopping routes that are not conventionally dominated, i.e., $\{\theta_i \in \Theta | ~\not\exists \theta_j \in \Theta : \theta_j \prec \theta_i\}$.
\end{defn}

A linear skyline consists of the subset of a conventional skyline that is optimal under \textit{all linear combinations} of the competing cost criteria \cite{Shekelyan2014}. Hence, in the scenario considered in this work a linear skyline is composed of combinations of stores that minimize the linear combination 
$\mathcal{F} = \delta_1 SC(\theta_i) + \delta_2 ST(\theta_i)$, 
with $\delta = (\delta_1, \delta_2)$ being a weight vector in $\mathbb{R}^2_{>0}$.
Note that we find the optimal solution for all such $\delta$, i.e., we do not require any weight vector to be provided beforehand. In the following, we remind the definition of $\delta$-dominance \cite{Shekelyan2014}, which determines when a combination of stores linearly dominates another one provided a particular $\delta$.

\begin{defn}[Linear dominance]
A shopping route $\theta_i$ is said to $\delta$-dominate another $\theta_j$ if and only if $\delta^T CV(\theta_i) < \delta^T CV(\theta_j)$.
\end{defn}

From the definition of linear dominance follows the definition of \textit{linear skyline} \cite{Shekelyan2014}.

\begin{defn}[Linear skyline]
\label{def: linear dominance}

Let $\Theta$ be a set of shopping routes. Let also $\Theta' = \{\theta_1, \ldots, \theta_K\} \subseteq \Theta$. Then, we say that $\Theta'$ linearly dominates a shopping route $\theta \in \Theta$ if and only if:
$$\big(\exists \theta' \in \Theta' : \theta' \prec \theta \big) 
\lor
\big(\forall \delta \in \mathbb{R}^2_{>0} \ \exists \theta' \in \Theta' : \delta^T CV(\theta') < \delta^T CV(\theta) \big)
$$
%
%
The maximal set of linearly non-dominated combinations of stores is referred to as \textit{linear skyline} and it can be seen as an \textit{ordered} set w.r.t. the first cost criteria.
\end{defn}

Testing the condition on the right hand side of Definition \ref{def: linear dominance} would require to try out every possible vector $\delta \in R^2_{>0}$, which would be computationally impractical. 
As such, in \cite{Shekelyan2014} the authors consider the problem from a different perspective, giving it an intuitive geometrical interpretation. More precisely,  the authors observe that a route $\theta$ is linearly dominated only if it lies \textit{above} the segment connecting any pair of shopping routes $\{\theta_q,\theta_j\}$ belonging to the linear skyline -- this fact is formally denoted by $\{\theta_q,\theta_j\} \prec_{L} \Theta$.
Then they show that it is possible to quickly test whether a route is linearly dominated or not (and thus determine if it can be added to the linear skyline).

Table~\ref{tab: notation} summarizes the main notation used throughout the rest of the paper.

\begin{table}[!ht]
  \centering
  \begin{tabular}{lr}
    \toprule
    \textbf{Symbol} & \textbf{Semantics} \\
    \midrule
    $\tau \in T$ & A store in the set of stores $T$\\
    $\theta_i = \langle \tau^i_1, \ldots, \tau^{i}_{n} \rangle$ & Shopping route\\
    $\tau^i_j \in R_i$ & The $j$-th store visited by $\theta_i$\\
    $v_\omega$ & Shopper's location\\
    $v_\sigma$ & Customer's delivery location\\
    $\lambda$ & Shopping list\\
    $ST(\theta_i)$ & Shopping time of route $\theta_i$\\
    $SC(\theta_i)$ & Shopping cost of route $\theta_i$\\
    $\theta_i \prec \theta_j$ & $\theta_i$ conventionally dominates $\theta_i$\\
    $\{\theta_i,\theta_j\} \prec_L \theta_q$ & $\theta_q$ linearly dominated by $\{\theta_i,\theta_j\}$\\
    $\theta^{SC}$ & Comb. of stores with minimum shopping cost\\
    $ST^U = ST(\theta^{SC})$ & Shopping time upper bound\\
    \bottomrule
  \end{tabular}
  \caption{{Main notation used throughout the paper.} 
  \label{tab: notation}}

\end{table}

\subsection{\NameProblem query's NP-hardness}
\label{app: np-hardness}

The \NameProblem query can be demonstrated to be NP-hard by showing that any instance of a 
Trip Planning Query (TPQ), a problem known to be NP-hard \cite{Li2005}, can be reduced to a PSD one.

Let us suppose to have a road network $G$, a set $C$ of categories of interest (COIs), and a set of points of interest (POIs) $P$ (also vertices in $G$), each belonging to some COI $c \in C$.
Given a starting location $v_s$, an ending location $v_d$, and a subset of COIs $C' \subseteq C$ provided by the user, a TPQ requires to compute the route with minimum cost (e.g., travel time) from $v_s$ to $v_d$ that visits exactly one POI from each COI in $C'$.

In the following we show how we can reduce any TPQ instance to a PSD one. $v_s$ can be trivially mapped to the shopper's location $v_{\omega}$, while $v_d$ can be trivially mapped to the customer's location $v_\sigma$. 
Next, we map the notion of subset of COIs to visit $C' \subseteq C$ with that of shopping list $\lambda$. 
Each COI $c \in C$ in TPQ can be mapped to the notion of a \textit{product} in PSD, with the POIs within that category representing the \textit{stores} selling that product.
Then, any subset $C' \subseteq C$ of COIs to visit can be expressed as a shopping list $\lambda$, where we require to buy \textit{exactly one unit} of each such fictitious products in $\lambda$.
Finally, recall that TPQ requires to compute the route minimizing the considered cost criterion (e.g., travel time), while the PSD query requires to find the set of linearly non-dominated routes w.r.t. shopping cost and shopping time. 
Considering that in TPQ there are no costs associated to POIs, we are free to 
impose that all the products have the same cost across all the stores in which they are sold, hence all the shopping routes satisfying $\lambda$ will have the same shopping cost. Therefore, the linear skyline will be a singleton containing the route with minimum shopping time, i.e., the optimal solution required by TPQ.


\subsection{Linear skylines, total order, and efficient insertions}
\label{app: efficient insertion check LS}

Let $ST$ be the cost criterion used to order the evaluation of shopping routes, let $\theta^{ST}$ be the route yielding minimum shopping time, let $LS$ be the skyline under construction -- initially $LS = \{\theta^{ST}\}$ --, and let $\theta$ be a shopping route considered for insertion. 
Shekelyan et al. \cite{Shekelyan2014} demonstrate that it suffices to verify whether $\theta$'s left \textit{neighbor} in the skyline, $LS_K$, i.e., the shopping route being the closest
to $\theta$ w.r.t. to $ST$, with $ST(LS_K) \leq ST(\theta)$\footnote{This corresponds to the last shopping route inserted into the skyline, or $\theta^{ST}$ in case no route was previously inserted.}, \textit{conventionally dominates} $\theta$. In other words, it suffices to verify whether $ST(LS_K) \leq ST(\theta)$.
If $ST(LS_K) > ST(\theta)$ then $\theta$ qualifies for insertion and it is necessary to verify if any route in the skyline becomes dominated due to the insertion of $\theta$.
First, it is necessary to verify whether $ST(LS_K) = ST(\theta)$: if that's the case, $LS_K$ is removed from the skyline as it is conventionally dominated by $\theta$.
Subsequently, thanks to the order in which linearly non-dominated shopping routes are discovered, it is sufficient to verify whether $\theta$'s left neighbor, $LS_K$, is linearly dominated by $LS_{K-1}$ ($LS_K$'s left neighbor) and $\theta$, i.e., verify if $\{LS_{K-1},\theta\} \prec_L LS_K$ is true, and remove $LS_K$ if this is the case. 
The procedure is iterated until $\theta$'s current left neighbor cannot be removed from the skyline or it represents the first route in the skyline, i.e., the shopping route with shopping time. Considering that linear skylines typically contain few elements, the overall cost of an insertion check can be assumed to be, on average, \textit{constant}.

\subsection{The \NameProblem query for the case of multiple customers}

\begin{figure} [htb]
\centering
\includegraphics[width=1\columnwidth]{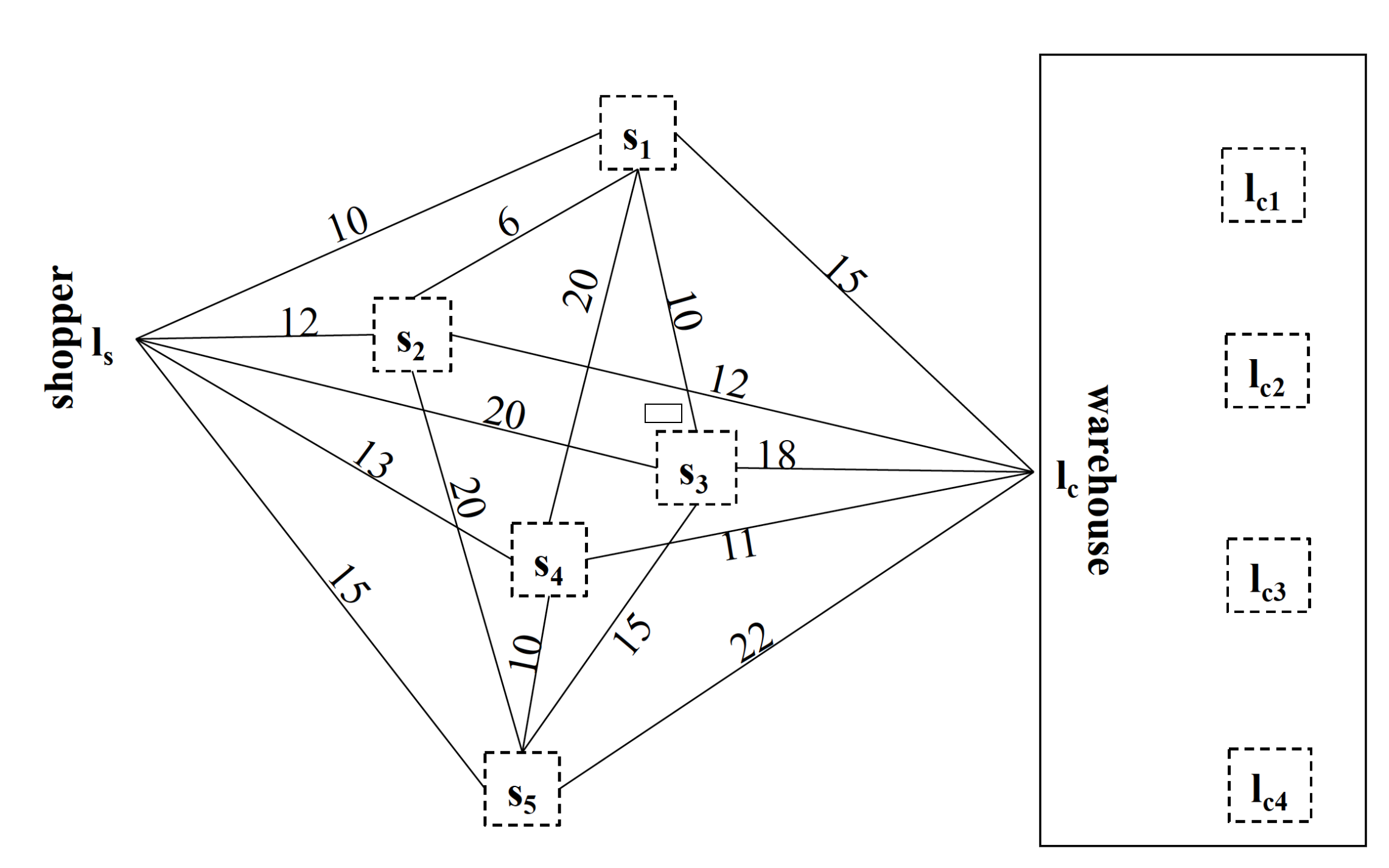}
\caption{Single shopper serving multiple customers when a warehouse is available.}
\label{fig:Multi_customer}
\end{figure}

A variation of the \NameProblem query as stated so far, is to have a single shopper that can serve several customers.  It can be trivially realized if one is allowed to establish a \textit{warehouse}. In this case, such a warehouse would effectively serve as a single artificial customer whose shopping list is the union of all actual customers' shopping lists. Observe then that the actual customers would be reached from the warehouse via paths that \emph{do not depend} on the shopping route chosen by the shopper to fulfill the single (combined) shopping list.  Figure~\ref{fig:Multi_customer} reuses the scenario discussed in Section~\ref{sec: intro} where there are four customers instead of a single one. Note that the road network connecting the warehouse and the customers is not shown, to reinforce that the ``delivery'' problem is immaterial and orthogonal to the \NameProblem's routing problem. 
Finally, one could possibly think of a more complex variant of the \NameProblem query where a shopper has to serve multiple customers 
\textit{without} having the possibility to rely on a warehouse. We leave such variant for future work.

%% file: 4_approach.tex
\section{Computing Solution for Personal Shopper's Dilemma Query}
\label{section:proposed approach}

In this section, we propose two approaches to solve the \NameProblem query. The first one, BSL-\NameProblem, is a baseline capable of computing \textit{optimal} linear skylines.
Given some shopping list, the strategy employed by BSL-\NameProblem evaluates in strict increasing order of shopping time shopping routes that fulfill the list, and orchestrate the construction of the skyline accordingly.
As shown in Section~\ref{app: np-hardness}, computing \NameProblem queries is NP-hard, and thus computing optimal linear skylines becomes unfeasible when the number of stores in a road network or the shopping list size becomes large.
We thus propose a second approach, APX-\NameProblem (Section \ref{sec: approximated approach}).
The idea behind APX-\NameProblem is to cluster stores spatially and then generate and combine shopping routes from such clusters, rather than from the entire set of stores, to reduce the number of shopping routes to possibly evaluate.
To this end APX-\NameProblem first superimposes a quad-tree over the stores within the considered geographical area.
Then, given a \NameProblem query APX-\NameProblem performs a depth-first search (DFS) of the quad-tree that is \textit{driven} by a \textit{scoring function}. Such function directs the search towards the tree leaves that are deemed the most ``\textit{promising}'' in terms of shopping time and shopping cost. The generation and expansion of shopping routes are thus conducted within single partitions, rather than on the entire set of stores, to considerably reduce the candidates' search space.

\subsection{A baseline approach: BSL-\NameProblem}
\label{sec: baseline approach}

The first approach we propose to solve the \NameProblem query is BSL-\NameProblem, a baseline capable of computing \textit{optimal} linear skylines. Before delving into its presentation, two important points of discussion concerning the strategy employed by BSL-\NameProblem to compute the linear skyline are: (i) our choice of \textit{imposing} a \textit{total order} over the evaluation of shopping routes -- from here on shopping routes under evaluation will be called \textit{candidate} routes -- and, (ii) the \textit{cost criterion} used to define such order.
For what concerns (i), \cite{Shekelyan2014} shows that having a total order over the candidates allows efficient insertions while constructing a linear skyline. Indeed, in order to determine if the candidate qualifies for insertion it suffices to verify whether a candidate is \textit{conventionally dominated} by the last shopping route inserted into the skyline. 


\begin{observation}
\label{obs: prop1 skylines}

Let us choose one of the two cost criteria as the primary cost. 
Let us also suppose that it is possible to compute the shopping route yielding the minimum possible value for the primary cost, which in turn establishes an upper bound for the second cost criterion, i.e., shopping routes having the second cost criterion larger than the upper bound are surely dominated by the associated route. 
Let us finally assume that linearly non-dominated shopping routes are evaluated \textit{in increasing order} w.r.t. the primary cost criterion. 
Then, it is sufficient to perform a single \textit{conventional domination} check to determine if a route qualifies for insertion into the linear skyline.
\end{observation}


For instance, 
Let $ST$ be the cost criterion used to order the evaluation of shopping routes, let $\theta^{ST}$ be the route yielding minimum shopping time, let $LS$ be the skyline under construction -- initially $LS = \{\theta^{ST}\}$ -- and let $\theta$ be a shopping route considered for insertion. 
Shekelyan et al. \cite{Shekelyan2014} demonstrate that it suffices to verify whether $\theta$'s left \textit{neighbor} in the skyline, $LS_K$, i.e., the shopping route being the closest
to $\theta$ w.r.t. to $ST$, with $ST(LS_K) \leq ST(\theta)$\footnote{This corresponds to the last shopping route inserted into the skyline, or $\theta^{ST}$ in case no route was previously inserted.}, \textit{conventionally dominates} $\theta$. In other words, it suffices to verify whether $ST(LS_K) \leq ST(\theta)$.
If $ST(LS_K) > ST(\theta)$ then $\theta$ qualifies for insertion and it is necessary to verify if any route in the skyline becomes dominated due to the insertion of $\theta$.
First, it is necessary to verify whether $ST(LS_K) = ST(\theta)$: if that's the case, $LS_K$ is removed from the skyline as it is conventionally dominated by $\theta$.
Subsequently, thanks to the order in which linearly non-dominated shopping routes are discovered, it is sufficient to verify whether $\theta$'s left neighbor, $LS_K$, is linearly dominated by $LS_{K-1}$ ($LS_K$'s left neighbor) and $\theta$, i.e., verify if $\{LS_{K-1},\theta\} \prec_L LS_K$ is true, and remove $LS_K$ if this is the case. 
The procedure is iterated until $\theta$'s current left neighbor cannot be removed from the skyline or it represents the first route in the skyline, i.e., the shopping route with shopping time. Considering that linear skylines typically contain few elements, the overall cost of an insertion check can be assumed to be, on average, \textit{constant}.

For the moment we assume said total order to be enforced by means of a min-priority queue $Q$, combined with some suitable generation scheme that progressively extends shopping routes according to the chosen cost criterion, and focuses on \textit{which} criterion should be used to define the evaluation order. Consider an evaluation strategy that evaluates candidates according to their increasing shopping cost. As it will be shown later on, computing the cost of a shopping route w.r.t. a shopping list can be done in constant time with the use of appropriate data structures. On the other hand, computing a candidate's shopping time always requires to compute the fastest path visiting its stores, i.e., it is necessary to solve an instance of the trip planning query, which is NP-hard. 
Thus, using shopping time rather than shopping cost allows for an evaluation strategy with greater pruning potential and thus smaller computational costs.
Now, note that evaluating in increasing order of shopping cost does not take into account the spatial information provided by a \NameProblem query (i.e., shopper's and customer's delivery locations), hence this strategy may end up evaluating shopping routes where the associated fastest paths are long and thus likely being dominated by faster routes. Overall, evaluating in increasing order of shopping time is less penalizing when incurring in dominated candidates, as the cost of computing their shopping cost is negligible.

At this point we can start focusing on the baseline's presentation. First, we briefly describe three pre-computed lookup tables BSL-\NameProblem uses to speed up key computations. Subsequently, we introduce the notion of skyline \textit{upper bounds}, and specify how they are computed and used within the baseline. We then proceed to introduce the set of pruning criteria and the generation scheme the baseline uses to evaluate candidates. Finally, we conclude by formally introducing  BSL-\NameProblem. 

We pre-compute three \textit{lookup tables}.
The first table keeps track of the stores where each product is sold, with the list of stores each product is associated with being ordered in increasing order of cost. This is similar to a typical text-based inverted list and allows us to determine in constant time a subset of stores yielding minimum shopping cost for any shopping list $\lambda$. 
The second table stores pairs $(product, store)$, with each pair being associated with the cost of the product at that store. Thus, for any shopping list $\lambda$ it is possible to compute in constant time if a combination of stores can fulfill it, as well as its shopping cost.
The third table stores the travel time of the fastest path connecting any pair of stores.

There are two important shopping routes BSL-\NameProblem uses to delimit the candidates' search space, i.e., the one yielding \textit{minimum shopping time}, $\theta^{ST}$, and the other yielding \textit{minimum shopping cost}, $\theta^{SC}$. The former provides the \textit{shopping cost upper bound}, $SC^U$, 
while the latter provides the \textit{shopping time upper bound}, $ST^U$.\\ 
\newline
\noindent \textbf{Computing $\theta^{SC}$.} Finding out a shopping route with minimum shopping cost requires to find out a subset of stores in $T$ where each product in $\lambda$ can be bought at minimum cost -- this can be done in constant time by using the \textit{first} pre-computed \textit{lookup table}.
Later on we show that during the candidates' evaluation it suffices to find the first shopping route with shopping cost equal to that of $\theta^{SC}$ to terminate the baseline's execution, as subsequent candidates are ensured to be longer (and thus dominated by such route).\\
\newline
\noindent \textbf{Computing $\theta^{ST}$.} To find out the shopping route with minimum shopping time requires to compute a variant of the trip planning query, i.e., it requires to find out the combination of stores that (i) yields the fastest path connecting the shopper's location, the stores, and the customer's delivery location, and (ii) that satisfies $\lambda$.
BSL-\NameProblem generates (and thus evaluates) a combination of stores in increasing order of shopping time. Consequently, the first candidate that fulfills $\lambda$ always corresponds to $\theta^{ST}$. We elaborate more on this shortly.\\ 

BSL-\NameProblem evaluates candidate shopping routes in increasing order of shopping time and constructs the linear skyline accordingly. To do so, it 
generates shopping routes according to the chosen order, coupled with a set of pruning criteria used to reduce the search space.
We start by presenting the first pruning criterion, a lemma that allows to update the shopping cost upper bound $SC^U$ as non-dominated routes are progressively inserted into the linear skyline.

\begin{lem}
\label{lemma: shopping cost update}

Let $Q$ be the min-priority queue used to order the evaluation of shopping routes in strict increasing order of shopping time. Let $\lambda$ be a shopping list. Let then $\theta$ be the last candidate popped from $Q$: if $\theta$ fulfills $\lambda$ \textit{and} qualifies for insertion into the linear skyline, then it is possible to set  $SC^U = SC(\theta)$.  

\begin{proof}
 The mechanism used to generate shopping routes in strict increasing order of shopping time ensures that any route $\theta'$ popped from $Q$ after $\theta$ yields $ST(\theta') \geq ST(\theta)$. Then, by virtue of the notion of linear skyline (Definition \ref{def: linear dominance}) $\theta'$ may qualify for insertion only if $SC(\theta') < SC(\theta) = SC^U$.
\end{proof}
\end{lem}

The above lemma allows to progressively enforce stricter upper bounds on shopping cost, and thus limit the candidates' space as linearly non-dominated solutions are found out.
Let us now introduce a lemma that allows to early terminate BSL-\NameProblem.

\begin{lem}
\label{lemma: algorithm termination 3}

Let $\theta$ be the last linearly non-dominated shopping route added to the linear skyline. If $SC(\theta) = SC(\theta^{SC})$ we are guaranteed to have found out all the linearly non-dominated shopping routes and the evaluation can terminate.

\begin{proof}
We know that shopping routes are popped from $Q$ in increasing order of shopping time. We also know that $\theta$ yields the minimum possible shopping cost. Consequently, any sequence $\theta'$ that fulfills $\lambda$ popped from $Q$ after $\theta$ necessarily has $ST(\theta') \geq ST(\theta)$ and $SC(\theta') \geq SC(\theta) = SC(\theta^{SC})$, i.e., Definition \ref{def: linear dominance} ensures that $\theta'$ is dominated. 
\end{proof}
\end{lem}

In other words, the above lemma guarantees that once a shopping route with minimum shopping cost $SC(\theta^{SC})$ and that satisfies $\lambda$ is found, the baseline can  terminate as it is not possible to find further non-dominated routes.

Let us now introduce a slightly modified notation for shopping routes that serves to relate them to the notion of \textit{ranked minimum detours}. Such notation will be used later on to illustrate the route generation scheme used by the baseline to evaluate candidates in increasing order of shopping time.

\begin{defn}[Shopping routes and minimum detours]
\label{def: notation shopping route detour}
We define a shopping route $\Theta = \langle \tau_{k_1}, \tau_{k_2}, \ldots, \tau_{k_{|\theta|}} \rangle$ to be the route where $\tau_{k_i} \in ST$ represents the $i$-th store visited by $\Theta$ and that yields the ${k_i}$-th minimum detour when added to the fastest path between the store that \textit{precedes} it in $\Theta$ (or the shopper's location $v_\omega$, if $i=1$) and the customer's delivery location $v_\sigma$.
\end{defn}

\begin{algorithm2e}[t]
\DontPrintSemicolon
\SetKwInOut{Input}{Input}
\SetKwInOut{Output}{Output}
\SetKwProg{myalg}{}{:}{end}
\SetKwFunction{csr}{{\sc{ComputeCSR_new}}}

{
    \Input{Road network $G$, set of stores $T \subseteq V$, shopper's current location $v_\omega$, customer's delivery location $v_\sigma$, shopping list $\lambda$, shopping route yielding minimum shopping cost $\theta^{SC}$.}
    \Output{Linear skyline $LS$.}
    
    
    $T \leftarrow$ \textsc{PruneStores}$(T, \lambda)$ \nllabel{alg:prune stores}\;
    $T \leftarrow$ \textsc{DijkstraMultipleTarget}$(v_{\omega}, T \cup \{v_\sigma\})$  \nllabel{alg:dijkstra shopper}\;
    $T \leftarrow$ \textsc{DijkstraMultipleTarget}$(v_{\sigma}, T)$  \nllabel{alg:dijkstra delivery}\;
    $LS \leftarrow \emptyset$ \nllabel{alg:init-start}\;
    $SC^U \leftarrow + \infty$\;
    $Q \leftarrow \{\langle \tau_{k_1 = 1} = \textsc{MinDetour}(v_\omega, v_\sigma, \emptyset, T) \rangle\}$ \nllabel{alg:init-end}\;
    
    \While{$Q \neq \emptyset$}
    {\nllabel{alg:while}
        $\theta = \langle \tau_{k_1}, \cdots, \tau_{k_{|\theta|}}\rangle, Q \leftarrow \textsc{Pop}(Q)$ \nllabel{alg:update queue}\;
        
            \If{\textsc{SatisfyList}$(\lambda, \theta)$}
            {\nllabel{alg:full route insertion}
                \If{$SC(\theta) < SC^U$}
                {\nllabel{alg:conventional domination}
                    $LS \leftarrow $\textsc{UpdateLS}$(LS,\theta)$\nllabel{alg:update skyline} \tcp*[r]{(Observation \ref{obs: prop1 skylines})}
                    $SC^U \leftarrow SC(\theta)$\nllabel{alg:update scu} \;
                }

                \lIf{$SC^U = SC(\theta^{SC})$}
                {\nllabel{alg: early terminate}
                    \Return LS
                }
            }
            
            $\theta^{s} = \langle \tau_{k_1}, \cdots, \tau_{k_{|\theta|}} \rangle \oplus  \textsc{MinDetour}(\tau_{k_{|\theta|}} , v_\sigma, \theta, T)$ \nllabel{alg: begin extend sequence}\;
            
            $\theta^{p} = \langle \tau_{k_1}, \cdots, \tau_{k_{|\theta|-1}} \rangle \oplus  \textsc{NextMinDetour}(\tau_{k_{|\theta|-1}}, v_\sigma, \theta, T)$\nllabel{alg: pred extend sequence}\;
                
            $Q \leftarrow \textsc{Push}(\{\theta^{s}, \theta^{p}\}, Q)$ \nllabel{alg: end extend sequence}
    }

    \Return $LS$ \nllabel{alg:final return}
}

\caption{ BSL-\NameProblem}
\label{alg:BSL_new}
\end{algorithm2e}

Algorithm \ref{alg:BSL_new} presents BSL-\NameProblem, along with the generation scheme used to evaluate shopping routes in strict increasing order of shopping time.
First, BSL-\NameProblem prunes from $T$ those stores that do not offer any of the products required in $\lambda$ (line \ref{alg:prune stores}, function \textsc{PruneStores}). This immediately reduces the candidates' search space.
Next, BSL-\NameProblem executes two single-source shortest path searches (lines \ref{alg:dijkstra shopper} and \ref{alg:dijkstra delivery}): the first one originates from the shopper's location $v_\omega$ and targets the set of stores $T$ as well as the delivery location $v_\sigma$. The second one originates from the delivery location $v_\sigma$ and targets the set of stores $T$. These searches return the travel times between $v_\omega$ and $v_\sigma$, and between any pair $(v_\omega, \tau)$ and $(\tau, v_\sigma)$, with $\tau \in T$.
Such travel times are then appropriately combined with those in the lookup table holding the travel times between any pair of stores to support efficient implementations of the functions \textsc{MinDetour} and \textsc{NextMinDetour} (discussed shortly). More specifically, for every store $\tau \in T$ we sum the travel time with any other store $\tau' \in T$ with that between $\tau'$ and $v_\sigma$. This yields a list of travel times which, once \textit{sorted}, allows finding quickly the store yielding the $k$-th minimum detour when added to the fastest path connecting $\tau$ and $v_\sigma$.

BSL-\NameProblem then goes on to set the initial state of several entities, namely, that of the linear skyline $LS$, the shopping time upper bound $ST^U$, and the priority queue $Q$, which initially holds the partial shopping route containing the store minimizing the detour w.r.t. the fastest path connecting $v_\omega$ and $v_\sigma$. 
Note that such store is found via the function \textsc{MinDetour}, which uses the information  computed previously.
Subsequently (while cycle, line \ref{alg:while}),
BSL-\NameProblem starts generating and evaluating shopping routes in increasing order of shopping time.
For each candidate $\theta$ popped from $Q$ (line \ref{alg:update queue}), BSL-\NameProblem first verifies whether $\theta$ fulfills $\lambda$ (line \ref{alg:full route insertion}).
If such condition holds, the baseline goes on to verify whether $SC(\theta) < SC^U$ (line \ref{alg:conventional domination}): if this condition does not hold, then $\theta$ is conventionally dominated (lemma \ref{lemma: shopping cost update}) and can be discarded. Otherwise, $\theta$ can be inserted into the skyline
according to the procedure outlined in observation \ref{app: efficient insertion check LS} (function \textsc{UpdateLS}, line \ref{alg:update skyline}), and $SC^U$ can be tightened (line \ref{alg:update scu}, by virtue of lemma \ref{lemma: shopping cost update}).
If $SC^U = SC(\theta^{SC})$ the algorithm immediately terminates (line \ref{alg: early terminate}), as it is not possible to find further linearly non-dominated shopping routes (by virtue of lemma \ref{lemma: algorithm termination 3}).

Lines \ref{alg: begin extend sequence}--\ref{alg: end extend sequence} represent the generation scheme used by BSL-\NameProblem to evaluate shopping routes in increasing order of shopping time, with $\oplus$ symbolizing the \textit{append} operation. For each shopping route $\theta$ popped from $Q$ the baseline generates two new shopping routes, namely, $\theta^s$ and $\theta^p$. $\theta^s$ is the shopping route generated by adding a store at the end of $\theta$ that (1) does not already appear in $\theta$ and that (2) \textit{minimizes} the detour distance when added to the fastest path between $\tau_{k_{|\theta|}}$ and the delivery location $v_\sigma$ -- such store is found out via the function \textsc{MinDetour}.
$\theta^p$ is the shopping route generated by replacing the last store visited by $\theta$, i.e., $\tau_{k_{|\theta|}}$, with the store yielding the $(k_{|\theta|} + 1)$-th minimum detour when added to the fastest path connecting $\tau_{k_{|\theta| - 1}}$ and $v_\sigma$ -- such store is found out via the function \textsc{NextMinDetour}. Similarly to $\theta^s$, note that we require the store found by \textsc{NextMinDetour} to not already appear in $\theta$. Observe that both $\theta^s$ and $\theta^p$ have shopping time greater or equal than $\theta$.

Finally, we can state two important features regarding BSL-\NameProblem:

\begin{thm}
\label{thm:gen_all_routes}
BSL-\NameProblem evaluates all possible shopping routes.  
\begin{proof}
\label{app: proof appendix:gen_all_routes}
We prove it by induction on the size of shopping routes. First we show that the baseline examines every shopping route with $|\theta|=1$. During the initialization phase BSL-\NameProblem initializes the min-priority queue with the shopping route yielding the minimum detour distance between $v_\omega$ and $v_\sigma$, i.e., $\theta = \langle \tau_{k_1 = 1} \rangle$ (line \ref{alg:init-end}). Then, every time some shopping route $\theta = \langle \tau_{k_1} \rangle$ is popped from $Q$, BSL-\NameProblem generates a shopping route $\theta^p$ (line \ref{alg: pred extend sequence}) where $\tau_{k_1}$ is replaced with $\tau_{k_1 + 1}$, i.e., the shop that yields the ${k_1}$-th minimum detour between $v_\omega$ and $v_\sigma$ is replaced with the one that yields the ${(k_1 + 1)}$-th minimum detour between $v_\omega$ and $v_\sigma$.
Let us now assume that BSL-\NameProblem generates all the shopping routes of length $n$. Then, for any route $\theta = \langle \tau_{k_1}, \ldots, \tau_{k_n} \rangle$ we know that the generation scheme generates a route $\theta^s = \langle \tau_{k_1}, \ldots, \tau_{k_n}, \tau_{k_{n+1} = 1} \rangle$ (line \ref{alg: begin extend sequence}), where the store $\tau_{k_{n+1} = 1}$ yields the minimum detour distance when added to the fastest path connecting $\tau_{k_n}$ and $v_\sigma$. Then, during subsequent iterations line \ref{alg: pred extend sequence} guarantees that $\theta^s$ allows to generate a whole set of shopping routes of length $n+1$ where the shop yielding the $k_{n+1}$-th minimum detour between $k_{n}$ and $v_\sigma$ is replaced with that yielding the $(k_{n+1} + 1)$-th minimum detour.

\end{proof}

\end{thm}

\begin{thm}
\label{thm:inc_routes}
BSL-\NameProblem evaluates shopping routes in strict increasing order of shopping time. 
\begin{proof}
\label{app: proof thm:inc_routes}
We prove it by contradiction. First, observe that the min-priority queue $Q$ ranks the shopping routes it holds in increasing order of shopping time. Hence, the only event that may violate the theorem's thesis is if a route popped from $Q$ has shopping time smaller than those popped (and thus evaluated) previously. 
Note, however, that such event is impossible, as the generation scheme used between lines \ref{alg: begin extend sequence}--\ref{alg: end extend sequence} ensures that the shopping routes $\theta^s$ and $\theta^p$ have shopping time always greater or equal than that of the popped route $\theta$ from which they are generated.\\
\end{proof}
\end{thm}



\subsubsection{BSL-\NameProblem's complexity}
\label{appendix: baseline complexity}
BSL-\NameProblem first filters out from $T$ stores that do not offer any product in $\lambda$. Thanks to the use of pre-computed lookup tables, the cost of such operation is $O(|\lambda| \cdot |T|)$.
Let us now denote by $T' \subseteq T$ the subset of stores that offer at least one product in $\lambda$. Let us also denote by $N = O\big(|T'|!)$ the overall number of candidate shopping routes to possibly evaluate, by $S$ the number of candidate routes $\theta^s$ generated at line \ref{alg: begin extend sequence}, and by $P$ the number of candidate routes $\theta^p$ generated at line \ref{alg: pred extend sequence}, with $1+S+P=N$.

The first major operations conducted by BSL-\NameProblem are the two single-source shortest path searches at lines \ref{alg:dijkstra shopper} and \ref{alg:dijkstra delivery}. Both searches have a common component cost of $O\big((|E|+|V|) \cdot log|V|\big)$, i.e., the cost inherent to conducting a Dijkstra search. Recall then that the second search is followed by the creation of a set of sorted lists, one per every store $\tau \in |T'|$, each holding the travel times that result from adding any other store in $|T'|$ in the fastest path connecting $\tau$ and $v_\omega$ (an information already available in the appropriate lookup table). Thus, creating and sorting such lists has cost $O(|T'| \cdot OC)$, where $OC$ is the sorting algorithm cost. 

 
Such operations are then followed by the baseline's time-dominant component, represented by the while loop (lines \ref{alg:while}--\ref{alg: end extend sequence}). The cost of executing such component can be expressed as $O\big(Nlog_2N + N2|T'| + S|\lambda| + |\lambda|\sum_{j=1}^P|\theta^p_j| + N \big)$.
The first term represents the cost associated with the use of a min-priority queue to order the evaluation of candidates.
The second term represents the cost of generating two candidate routes from each candidate popped from the queue (i.e., the cost of executing \textsc{MinDetour} and \textsc{NextMinDetour}). By using the aforementioned sorted lists, finding the $k$-th minimum detour from any store to the delivery location has cost $O(|T'|)$, which in turn yields $O(N2|T'|)$.
The third term represents the cost of computing the shopping cost of candidates $\theta^s$ generated at line \ref{alg: begin extend sequence} from some candidate $\theta$. Recall that each product in $\lambda$ shall be bought at the store selling it for the lowest price among those in $\theta$. Hence, computing the shopping cost of a candidate $\theta^s$ can be done in $O(|\lambda|)$ by (1) using the lookup table providing the selling price of any product in any store and (2) by keeping track of the minimum price each product in $\lambda$ is bought among $\theta$'s stores. The fourth term represents the cost of computing the shopping cost of candidates $\theta^p$ generated at line \ref{alg: pred extend sequence}. Recall that such operation requires to replace the last store in some $\theta$ with the one yielding the next minimum detour. Now, observe that some of the products in $\lambda$ may be bought at the store being replaced, hence in the worst case its replacement requires to find out for each product in $\lambda$ which store among those in $\theta^p$ sells it at minimum price.
Finally, the fifth term represents the cost needed to check if candidate routes qualify for insertion into the linear skyline.

\subsection{An approximated approach: APX-\NameProblem}
\label{sec: approximated approach}

The baseline's major drawback lies in its necessity to possibly evaluate a number of shopping routes that is \textit{factorial} in the number of stores, thus limiting its scalability and applicability to real-world scenarios.
In order to overcome that, we propose APX-\NameProblem, an approach that trades the optimality of linear skylines for a greatly reduced number of candidates to evaluate.
The key idea behind APX-\NameProblem is to consider stores at a \textit{coarser granularity}, i.e., \textit{partitions} of stores rather than individual stores to generate -- and thus evaluate -- shopping routes from those partitions, rather than the whole set of stores, that look the most ``promising.'' 

To partition the stores of a road network APX-\NameProblem superimposes a point-region (PR) quad-tree \cite{prquadtree} 
over the minimum bounding rectangle (MBR) enclosing the stores. This corresponds to the \textit{root} quadrant of the quad-tree.
The PR quad-tree is then constructed by recursively splitting each quadrant having a number of elements (stores) larger than a given capacity threshold in four sub-quadrants. Quadrants that are split during the construction process are the quad-tree \textit{intermediate} nodes, while unsplit ones make up the \textit{leaves}. 
Given that it relies only on the spatial location of the stores, APX-\NameProblem can pre-compute a quad-tree over the stores of a road-network, along with information concerning travel time \textit{between partitions} and \textit{statistics} on the products each quadrant (at any level of the tree) holds. 
Such statistics are subsequently used to drive the generation and evaluation of candidate routes, i.e., to decide which quadrants are the most \textit{promising} w.r.t. shopping time and shopping cost.

We first define the notions of \textit{travel time} between \textit{quadrants} and \textit{travel time} between a \textit{vertex} and some \textit{quadrant}, as they are key to the evaluation strategy employed by APX-\NameProblem.

\begin{defn}[Travel time between quadrants]
\label{def: tt partitions}

Let $Q$ be a PR quad-tree superimposed over the stores $T \subseteq V$. Let $P_i$ and $P_j$ be two quadrants of $Q$.
Then, the travel time between $P_i$ and $P_j$, denoted by $mTT(P_i,P_j)$, is defined by $mTT(\tau^{i}_{k}, \tau^{j}_{l})$, with $\tau^{i}_{k} \in P_i$,  $\tau^{j}_{l} \in P_j$, and $\nexists (\tau^{i}_{m}, \tau^{j}_{n})$, with $\tau^{i}_{m} \in P_i$, $\tau^{j}_{n} \in P_j$, such that $mTT(\tau^{i}_{m}, \tau^{j}_{n}) < mTT(\tau^{i}_{k}, \tau^{j}_{l})$.
\label{def:part_TT1}
\end{defn}

\begin{defn}[Travel time between a vertex and a quadrant]
\label{def: tt vertex partition}

Let $Q$ be a PR quad-tree superimposed over the stores $T \subseteq V$. Let $v \in V$ be some vertex and $P \in Q$ be some quadrant. Then, the travel time between $v$ and $P_j$, denoted by $mTT(v, P)$, is defined by $mTT(v, \tau^{j})$, with $\tau^{j} \in P$, and $\nexists (v, \tau^{j}_{n})$, with $\tau^{j}_{n} \in P_j$, such that $mTT(v, \tau^{j}_{n}) < mTT(v, \tau^{j}_{l})$. 
\label{def:part_TT2}
\end{defn}

Observe that Definition \ref{def: tt partitions} requires to find out the pair of stores, each belonging to one of the considered quadrants, yielding minimum travel time, while Definition \ref{def: tt vertex partition} requires to find out the store within a quadrant that \textit{minimizes} travel time w.r.t. some given vertex.
Finally note that travel times between partitions can be pre-computed.
Both definitions allow APX-\NameProblem to enforce upper bounds on shopping time when generating shopping routes from different partitions -- to be discussed further shortly.

With a PR quad-tree in place, APX-\NameProblem can then proceed to process \NameProblem queries. Its evaluation strategy relies on a \textit{depth-first} search (DFS) of the quad-tree driven by a \textit{scoring} function we introduce below. 
Given the set of shopping routes under construction (initially empty), such function estimates how ``good'' each quad-tree quadrant, be it an intermediate node or a leaf, is in terms of shopping time and shopping cost w.r.t. the characteristics of a \NameProblem query and the shopping routes generated so far, thus driving the generation and evaluation of shopping routes towards quadrants that are deemed the ``most promising''.

\begin{defn}[Quad-tree quadrant scoring function]
Let us consider a shopping list $\lambda$, a PR quad-tree $Q$ superimposed over $T \subseteq V$, and a quadrant $P_i \in Q$ (either intermediate node or leaf) that needs to be scored w.r.t. $\lambda$.
Let us assume that $P_i$ must be reached from either the shopper's location $v_\omega$ or some other quadrant $P \in Q$.
Let us further consider that $P_i$ contains $m$ of the products specified in $\lambda$; let us denote by $\lambda^{P_i}$ the subset of such products, and by $\lambda^{P_i}_k$ the average cost of the $k$-th product within $\lambda^{P_i}$ among $P_i$'s stores.

Let us finally denote the minimum travel time needed to depart from $v_\omega$, visit one or more stores in $P_i$, and finally reach the customer's delivery location $v_\sigma$ by $ST^{v_\omega}_{P_i} = mTT(v_\omega, P_i) + mTT(v_\sigma, P_i)$. Analogously, departing from any $P \in Q$ yields $ST^P_{P_i} = mTT(P, P_i) + mTT(v_\sigma, P_i)$. 
For the sake of simplicity, in this context we denote either $v_\omega$ or $P$ by $x$. Let us finally denote by $maxPrice$ the price of the most expensive product in any store from $\lambda$.
Then, we define the score of $P_i$ w.r.t. $x$ and $\lambda$ as follows: 
\begin{equation}
\label{eq: scoring function}
\mbox{\(
F(x, \lambda, P_i) =
\begin{cases} 
        
       \frac{ ST^x_{P_i}}{ ST^U} + \frac{\sum_{k=1}^{|\lambda^{P_i}|} \big(\lambda^{P_i}_k / maxPrice \big)}{m} & m > 0 \\
      +\infty & m = 0
      
\end{cases}
\)} %
\end{equation}
We normalize each of the two terms on the right hand side of the equation to give the same importance to shopping time and shopping cost. Specifically, $ST^x_{P_i}$ is normalized w.r.t. the shopping time upper bound $ST^U$, while the average cost of each product is normalized w.r.t. the cost of the most expensive product available. 
Consequently, assuming that $m >0$ we have $\forall(x,\lambda,P_i), F(v,\lambda,P_i) \in [0,2]$.
\end{defn}

Intuitively, the lower the score, the more ``promising'' a quadrant is. Observe also that the scoring function attempts to balance the importance given to shopping time and shopping cost. We thus expect APX-\NameProblem to generate shopping routes with shopping time and cost having the tendency to distribute more toward the centers of the intervals in the corresponding optimal skyline. 
In fact, the experimental results shown in Section \ref{section:experimental evaluation} confirm that a scoring function with such a feature makes APX-\NameProblem \textit{robust} to variations in store and products' price \textit{distributions}.

\begin{algorithm2e}[htb]
\DontPrintSemicolon
\SetKwInOut{Input}{Input}
\SetKwInOut{Output}{Output}
\SetKwProg{myalg}{}{:}{end}
\SetKwFunction{csr}{{\sc{ComputeCSR}}}

{
    \Input{road network $G=(V,E)$, set of stores $T \subseteq V$, quad-tree $Q$ superimposed over the MBR enclosing the elements in $T$, shopper's location $v_\omega$, delivery location $v_\sigma$, shopping list $\lambda$, shopping time upper bound $ST^U$.}
    \Output{approximated linear skyline $LS$.}
    
    \BlankLine
    $Q,T \leftarrow$ \textsc{PruneStores}$(T, \lambda)$ \nllabel{alg:apx_prune stores}\;
    $T \leftarrow$ \textsc{DijkstraMultipleTarget}$(v_{\omega}, T \cup \{v_\sigma\})$  \nllabel{alg:apx_dijkstra shopper}\;
    $T \leftarrow$ \textsc{DijkstraMultipleTarget}$(v_{\sigma}, T)$  \nllabel{alg:apx_dijkstra delivery}\;
    $P \leftarrow \textsc{GetRoot(Q)}$ \nllabel{alg:init-start-APX}\;
    $LS  \leftarrow$ \textsc{Explore}$(P, Q, \lambda, v_\sigma, \emptyset, \emptyset, ST^U)$
    \nllabel{apx:exp}\;



    \Return $LS$ \nllabel{PSD:Ret}
}

\caption{APX-PSD}
\label{alg:PSD}
\end{algorithm2e}

\begin{algorithm2e}[htb]
\DontPrintSemicolon
\SetKwInOut{Input}{Input}
\SetKwInOut{Output}{Output}
\SetKwProg{myalg}{}{:}{end}
\SetKwFunction{ex}{{\sc{Explore}}}
\Input{current partition $P$, quad-tree superimposed over $G$, $Q$, shopping list $\lambda$, customer's delivery location $v_\sigma$, linear skyline $LS$, set of partial shopping routes $PR$, shopping time upper bound $ST^U$.}
\Output{set of partial shopping routes $PR$ (updated), linear skyline $LS$ (updated), shopping list $\lambda$ (propagated).}
   
\BlankLine
$\lambda' \leftarrow$ \textsc{GetMissingProducts}$(\lambda, PR)$\nllabel{line: get missing products}\;
\lIf{$\lambda' = \emptyset$}
{
    \Return $PR, LS, \lambda$
}
\If{\textsc{IsLeaf}$(P,Q)$}
 {\nllabel{line: isleaf}
    $PR, LS \leftarrow$ \textsc{ComputePartitionRoutes}$(\lambda', P, PR, LS)$
    \nllabel{alg:combo1}\;
 \nllabel{alg:term}
 }
\Else
{\nllabel{alg:leafsearch1}
    $src \leftarrow$ \textsc{GetStart}$(PR)$\nllabel{line: getsrc}\;
    $Z = \{P_1,P_2,P_3,P_4\} \leftarrow$ \textsc{Score} $(Q, P, \lambda, src, v_\sigma, PR, ST^U)$\;
    \While{$Z \neq \emptyset$}
    {
        $z \leftarrow \textsc{GetTopScore}(Z)$\;
        $PR, LS, \lambda \leftarrow$ \textsc{Explore}$(z, Q, \lambda, LS, PR, ST^U)$\nllabel{line: recursive score}\;
        $Z \leftarrow Z \setminus \{z\}$, $src \leftarrow$ \textsc{GetStart}$(PR)$\;
        $Z \leftarrow$ \textsc{ReScore} $(Q, Z, \lambda, src, v_\sigma, PR, ST^U)$\nllabel{line: rescore}\;
    }
}

\Return $PR, LS, \lambda$\;

\caption{\large{\textsc{\textbf{Explore}}}}
\label{alg:Exp}
\end{algorithm2e}

At this point we are ready to introduce APX-\NameProblem. 
Algorithms \ref{alg:PSD} and \ref{alg:Exp} present the pseudo-code behind our approach.
Algorithm \ref{alg:PSD} starts by performing several  preliminary operations. First, it removes from $T$ stores that do not offer any of the products in $\lambda$, and updates $T$ and $Q$ accordingly (line \ref{alg:apx_prune stores}). Next, it performs two single-source shortest path searches at lines \ref{alg:apx_dijkstra shopper}--\ref{alg:apx_dijkstra delivery} that, analogously to those in BSL-\NameProblem, compute the fastest paths between the shopper's and customer's delivery locations and the stores in $T$. 
The algorithm then initiates to recursively perform a depth-first visit of the quad-tree, with the goal of constructing shopping routes from the tree's leaves that look the ``most promising''
and update the linear skyline accordingly.
Such operations are implemented by the function \textsc{Explore}, invoked at line \ref{apx:exp}. Algorithm \ref{alg:Exp} provide the details.

\textsc{Explore} first determines the set of products that cannot be bought from the routes currently stored in $PR$ and stores such set in $\lambda'$ (line \ref{line: get missing products}, function \textsc{GetMissingProducts}).
Note that if $\lambda' = \emptyset$ then \textsc{Explore} can \textit{terminate}, as this implies that the depth first search conducted within $Q$ already found out shopping routes that can satisfy $\lambda$ and inserted them into $LS$. 
\textsc{Explore} then verifies if the currently considered quadrant $P$ (initially the quad-tree root) is a \textit{leaf} or not (line \ref{line: isleaf}). 
If $P$ is a leaf, then the function \textsc{ComputePartitionRoutes} is executed (line \ref{alg:combo1}). Such function first counts the number of products in $\lambda'$ that can be bought from $P$'s stores -- let us suppose $m$ products. 
Subsequently, the function generates the set of (partial) shopping routes from $P$'s stores, where \textit{each} such route allows to buy \textit{exactly} those $m$ products (stores that do not offer any of the products in $\lambda'$ are ignored). 
\textsc{ComputePartitionRoutes} then goes on to compute the \textit{Cartesian product} between the set of routes currently stored in $PR$ and that computed from $P$. The result then becomes the new $PR$'s content. 
Each route in $PR$ is subsequently checked to verify if its shopping time is above $ST^U$ -- in such case the route is removed from $PR$. Finally, \textsc{ComputePartitionRoutes} verifies if the surviving routes buy all the products in $\lambda$ (i.e., when $\lambda' = \emptyset$) and, if so, attempts to insert them into $LS$.

If $P$ is an intermediate node of $Q$ then \textsc{Explore} first proceeds to determine in which partition the routes currently stored in $PR$ terminate and set $src$ accordingly (line \ref{line: getsrc}, function \textsc{GetStart}). Then, \textsc{Explore} scores $P$'s four sub-quadrants (line \ref{alg:leafsearch1}, function \textsc{Score}) and finally recursively invokes itself by considering such sub-quadrants in ascending order of score (line \ref{line: recursive score}). Note that the function \textsc{ReScore} (line \ref{line: rescore}) takes advantage of any update in $PR$ to update the scores of partitions still within $Z$, and thus better direct the evaluation of candidate routes.



\subsubsection{APX-\NameProblem's complexity}
\label{appendix: appx complexity}

APX-\NameProblem first requires to pre-compute a quadtree, an operation that has cost $O(|T|)$.
The two single-source shortest searches (lines \ref{alg:dijkstra shopper} and \ref{alg:apx_dijkstra delivery}) operate on $G=(V,E)$ and have cost $O\big((|E|+|V|) \cdot log|V|\big)$ each.
Let us now focus on the recursive execution of \textsc{Explore}. 
The cost of such execution is $O\big ((|V^Q| + |E^Q|) + N' + N + N|T'||\lambda| + NlogN)$.

The first term is due to the DFS being conducted over $Q$: if $V^Q$ and $E^Q$ denote, respectively, the nodes and edges in $Q$, then performing a DFS has cost $O(|V^Q| + |E^Q|)$.
The second term, $N'$, represents the cost of generating shopping routes from $Q$'s leaves by \textsc{ComputePartitionRoutes}: if $L$ denotes the set of $Q$'s leaves, then $N'= \sum_{i=1}^{|L|}|L_i|!$.
The third term represents the cost of generating (partial) shopping routes by means of the Cartesian products conducted within \textsc{ComputePartitionRoutes}, with $N = \prod_{i=1}^{|L|}|L_i|!$.
The fourth term represents the cost of finding the store, within each of the $N$ shopping routes, from which a product in $\lambda$ can be bought at minimum price. Finally, the fifth term represents the cost of generating the final linear skyline, i.e., $O\big(NlogN)$.


Overall, APX-\NameProblem has to evaluate much less candidate routes (if $L$ denotes the set of $Q$'s leaves, then $O(\prod_{i=1}^{|L|}|L_i|!)$) than BSL-\NameProblem ($O(|T|!)$), thanks to the spatial partitioning imposed over the stores and how such partitioning is used to generate and evaluate candidate shopping routes. On the other hand, this limits the evaluation to a restricted subset of candidate routes, which explains APX-\NameProblem's expected sub-optimality.

%% file: 5_experiments.tex
\section{Experimental Evaluation}
\label{section:experimental evaluation}

In order to evaluate the APX-PSD approach, we used datasets containing the road networks and POI locations for Amsterdam, Oslo and Berlin \cite{Dataset}. Since those POIs do not include actual stores we used the locations of gas stations and pharmacies (which are typically scattered throughout a city) as proxies for the locations of stores.

\begin{figure}[htb]
  \centering
  
    \includegraphics[width=0.36\columnwidth]{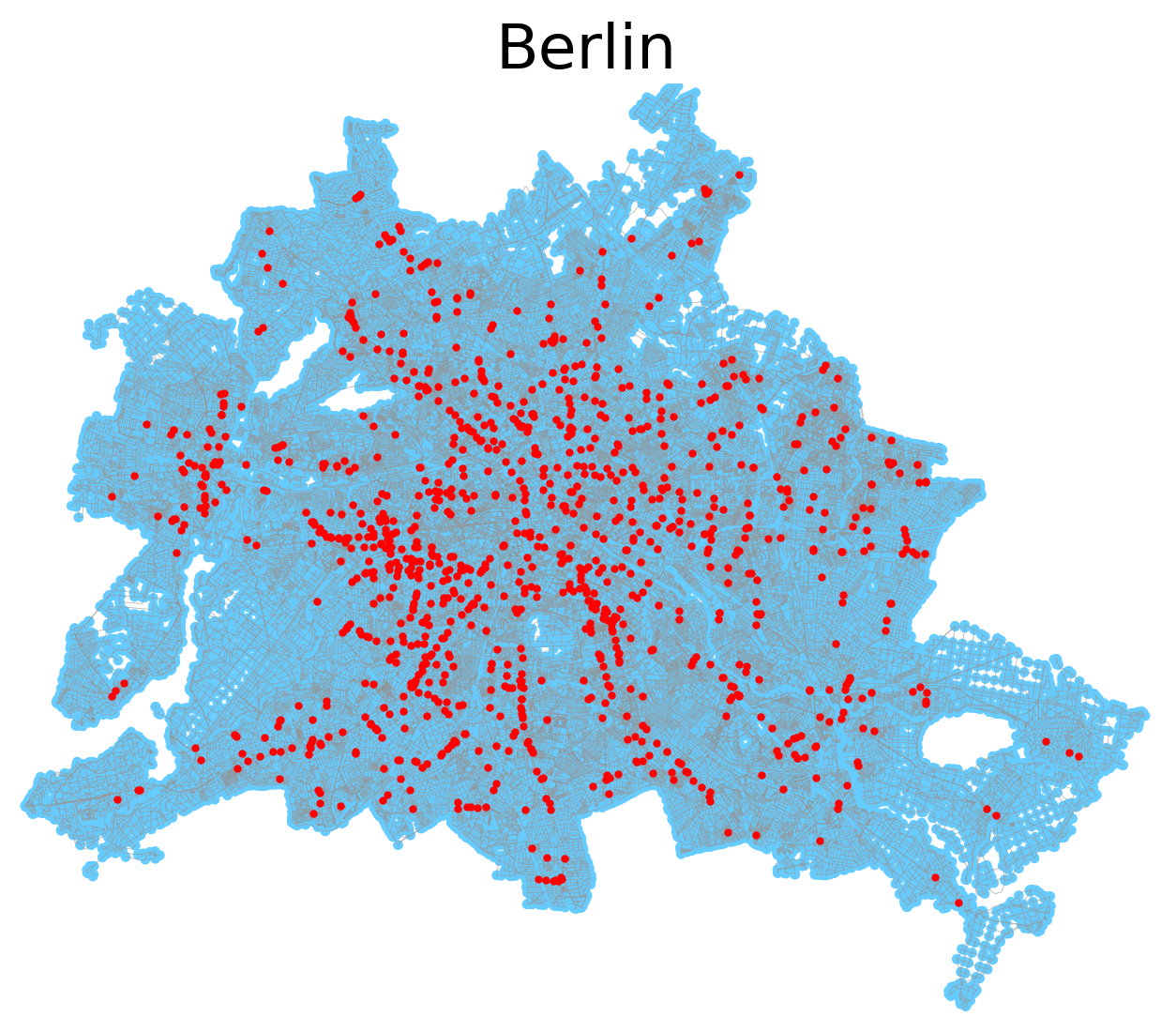}
    \includegraphics[width=0.25\columnwidth]{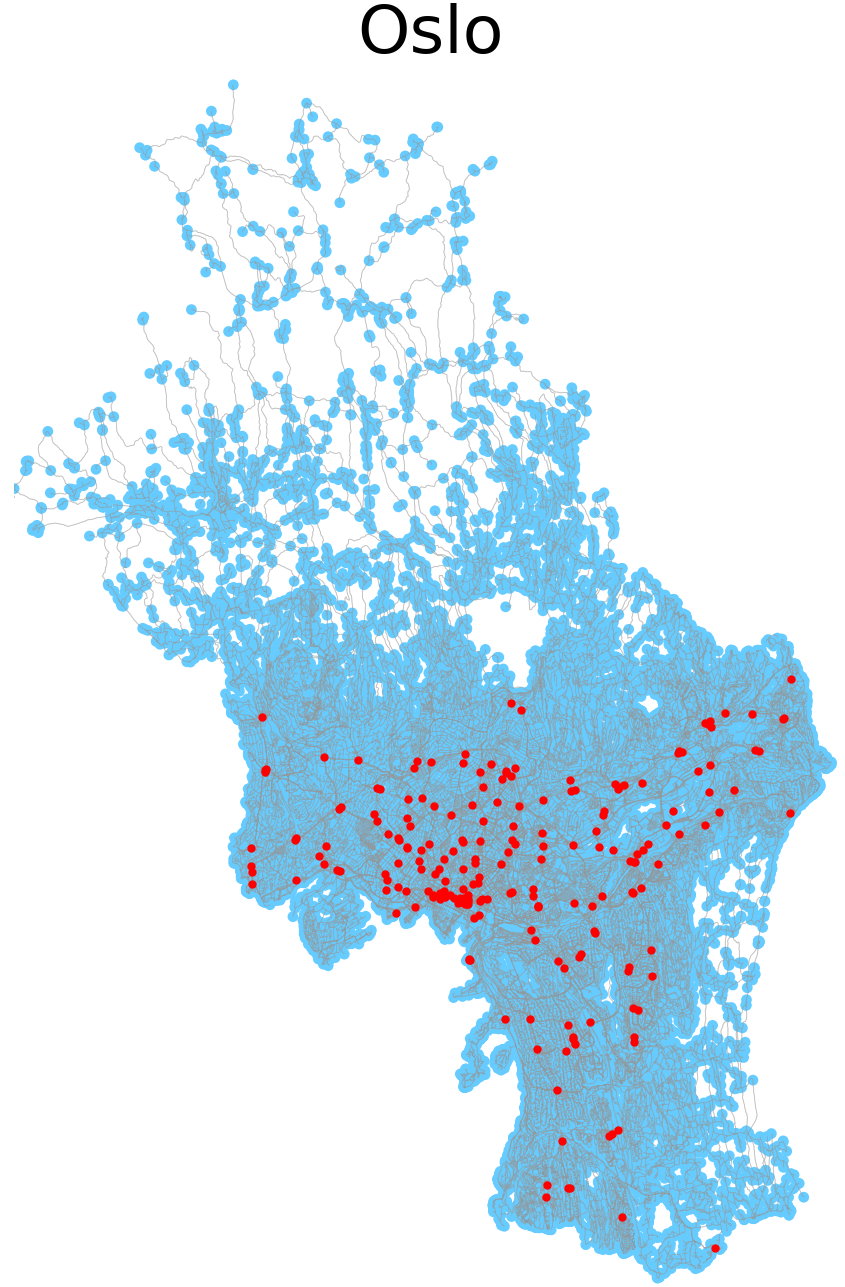}    \includegraphics[width=0.37\columnwidth]{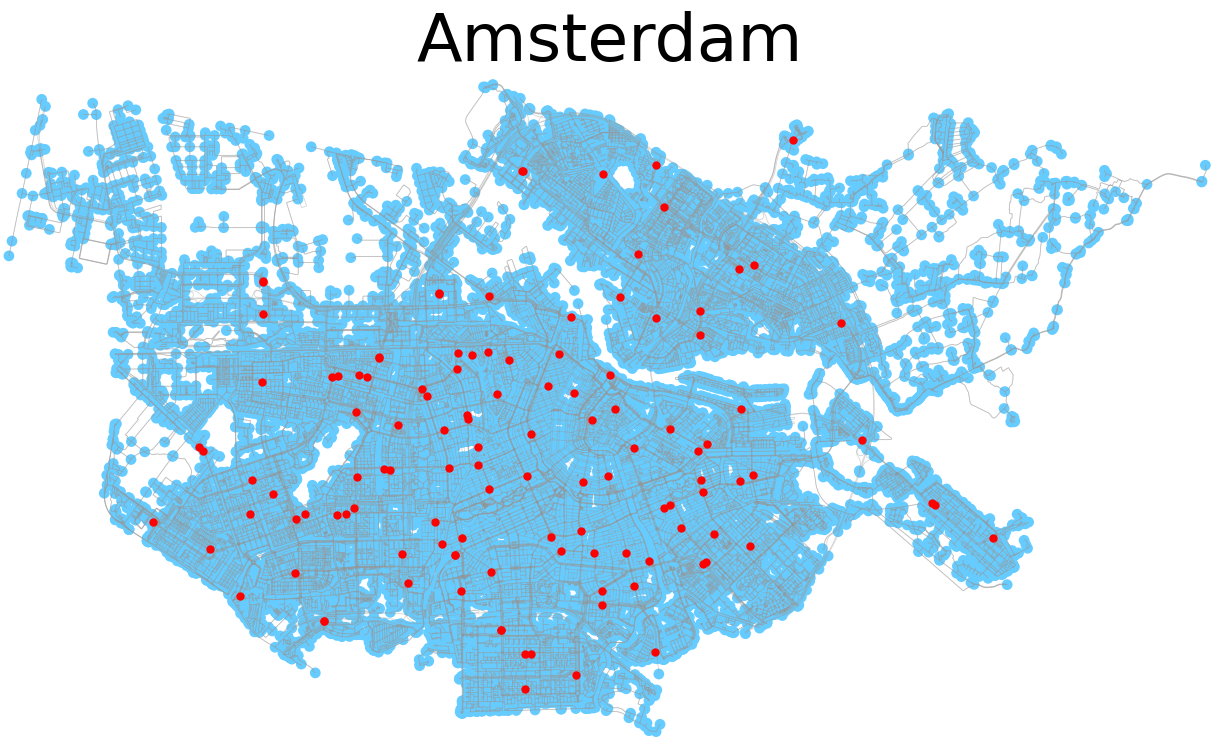}

  \caption{{Distribution of stores in each city's road network. (Cities not shown to scale).}}
  \label{fig:maps}
\end{figure}

Figure~\ref{fig:maps} illustrates the store locations in Berlin, Oslo and Amsterdam as used in our experiments. Recall that pharmacies and gas stations were used as proxy for stores, due to (1) the absence of information about real stores in these cities, and (2) it is realistic to assume that pharmacies and gas stations are scattered over a city just like stores would be.
The total number of vertices, edges and stores for all of the three maps are shown in Table~\ref{tab: datasets}.

\hspace{1pt}
\begin{table}[htb]
  \centering
  \begin{tabular}{lccc}
    \toprule
    & \textbf{Amsterdam} & \textbf{Oslo} & \textbf{Berlin}\\
    \midrule
    \vspace{3pt}
    Vertices & 106600 & 305175 & 428769\\
    \vspace{3pt}
    Edges & 130091 & 330633 & 504229\\
    Stores & 100 & 207 & 768\\
    \bottomrule
  \end{tabular}
  \caption{{Metadata for datasets.}}
  \label{tab: datasets}
\end{table}

The parameters considered for the experimental evaluation are: 
(1) store cardinality (i.e., number of stores in a network),
(2) distribution of a products' cost in the stores,
(3) spatial distribution of differently sized stores
(4) size of shopping lists, and
(5) capacity of the quad-tree leaves.
%

Given that the networks are fixed, varying the store cardinality models the density of stores in a city.
We also looked into the case when prices increase or decrease w.r.t. the distance of the store to the city's center.
We considered three different sizes of stores, i.e.,  small, medium and large, according to the percentage of the total number of products they hold, i.e., 25\%, 50\% and 75\%, respectively, out of a total of 1,000 products.
Note that we chose to not have any store selling all the products to avoid a possibly trivial solution involving a single store.
Even though stores are by default distributed randomly throughout the network, regardless of their sizes, we also investigated the effect of increasing or decreasing the size of stores according to their proximity to the city's center.
The values used for each parameter are shown in Table~\ref{tab:defaultVal}.

\hspace{1pt}
\begin{table}[htb]
  \begin{center}
    \begin{tabular}{lc} 
     \toprule
      \textbf{Parameter} & \textbf{Default Value}\\
        \midrule
       \vspace{3pt}
       Store  Cardinality & 10, \textbf{25}, 50\\
       \vspace{3pt}
       Product Cost Distribution & Rising, \textbf{Normal}, Declining \\
       \vspace{3pt}
       Store Size Spatial Distribution & Increasing, \textbf{Random}, Decreasing\\
       \vspace{3pt}
       Shopping list size & 5, \textbf{10}, 15\\
       \vspace{3pt}
       quad-tree leaf capacity & 4, \textbf{8}, 16\\
      \bottomrule
    \end{tabular}
    \caption{{Default Parameter Values.}}
    \label{tab:defaultVal}
  \end{center}
\end{table}

We attempted to emulate a realistic scenario in terms of the distribution of the cost of products as well as of placement of differently sized stores. 

We thus divide the maps into \textit{three concentric rings} w.r.t. the city center. The inner, middle and outer rings will host small, medium and large stores, respectively, when the spatial distribution of stores is set to ``Increasing''.  Conversely, when that parameter is set to ``Decreasing'' the inner, middle and outer rings segment will host large, medium and small stores, respectively. 

Similarly, when the product cost distribution parameter is set to ``Rising" (Declining) products in stores farther (closer) from the city's center are more expensive. 
The store closest (farther) to the city centre will sell it at minimum (maximum) price and the one farthest will sell it at the maximum (minimum) price. 
All the stores in between will sell the product at a price proportional to the distance between those two stores.
For ``Normal'' distribution we first find a mean price for each product following a U(5,15) distribution. Once we have the mean price for a product, we assign prices of that product to different stores following a normal distribution with its mean price and a standard deviation of $2$.  As discussed in Section \ref{sec: approximated approach}, APX-PSD's evaluation strategy relies on partitions of stores. Thus we observe the effects of the leaf capacity on the APX-\NameProblem by varying the leaf capacity to 4, 8 and 16.

Finally, we test each value each considered parameter  can assume by conducting $100$ individual experiments, and report the average optimality gap, coverage gap and processing time. 
In each experiment we randomly select the shopper's and customer's delivery locations, as well as randomly generate a shopping list of the required size. 
Furthermore, we randomly select the required number of stores ($25$ stores in default setting) from those available in the networks. 
All the experiments were conducted on a virtual machine with an Intel(R)-Xeon(R) CPUs running at $2.30$GHz and with $264$GB of RAM. 

\subsection{Evaluation metrics}

We evaluate our APX-\NameProblem by measuring the quality of the results (effectiveness) as well as the efficiency of it. In this section, we discuss the metrics we use to measure such effectiveness and efficiency.

\subsubsection{Effectiveness}
Comparing an optimal linear skyline (opt-LS) to an approximated one (apx-LS) requires comparing two aspects of skylines: optimality and coverage. 
For that we propose the two measures presented next.

\begin{figure} [htb]
   \centering
   \includegraphics[width=0.5\columnwidth]{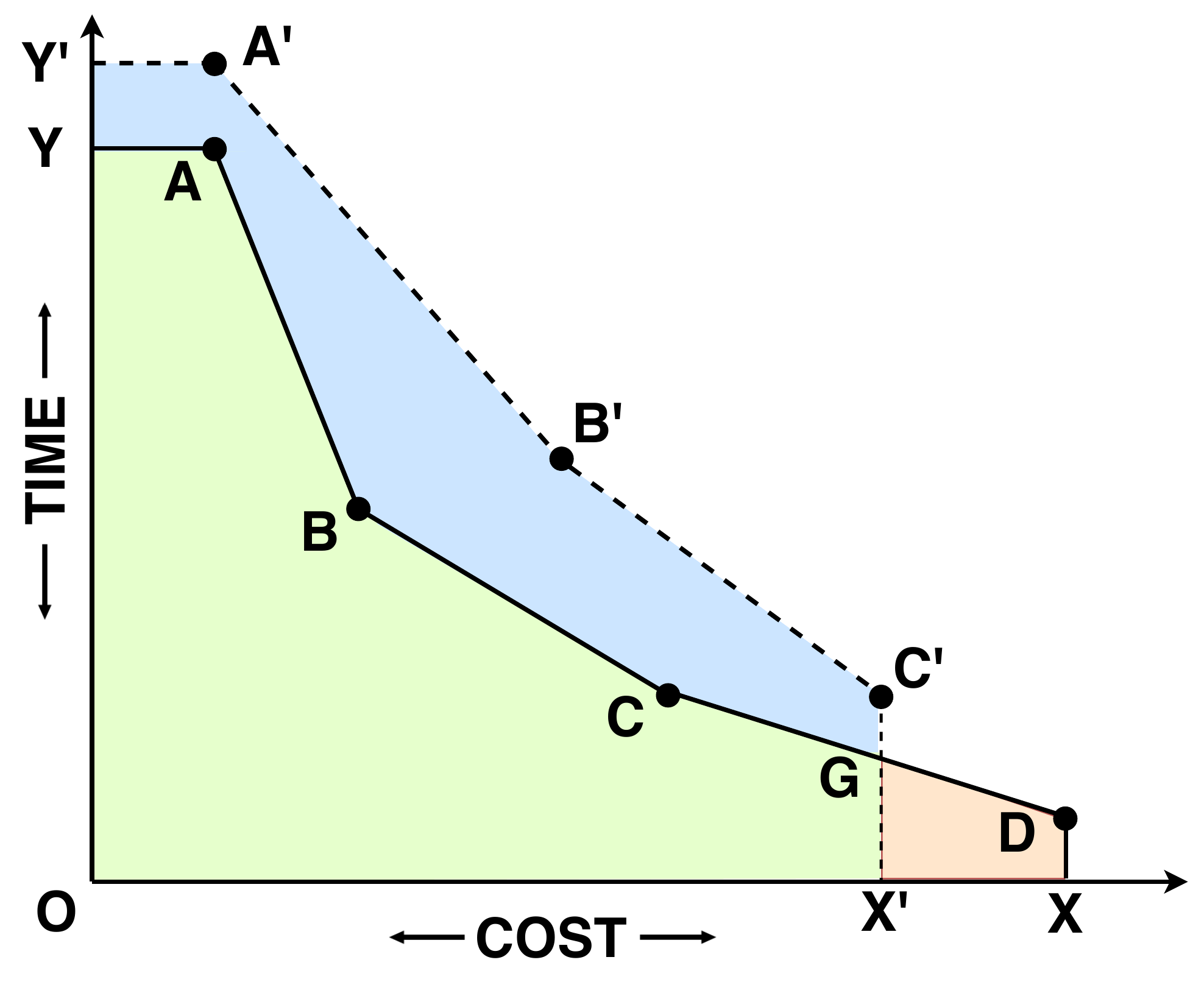}
   \caption{{Area coverage by the approximate and optimal linear skyline.}}
   \label{fig:RefLine}
\end{figure}

Consider Figure~\ref{fig:RefLine} where opt-LS = $\{A, B, C, D\}$ and apx-LS = $\{A', B', C'\}$. 
The area given by the polygon $OYABCDXO$ (shaded in green plus orange) represents the area $A_{opt}$ not dominated by opt-LS, and similarly the polygon $OY'A'B'C'X'O$ (shaded in blue \emph{plus} the one shaded in green) denotes the area $A_{apx}$ not dominated by apx-LS. The smaller the difference between $A_{apx}$ and $A_{opt}$ the better, but in order to make the right comparison we need to consider only the portion of $A_{apx}$ that intersects with $A_{opt}$, which in the case of this example is given by the polygon $OYABCGX'O$, which we denote by $A_{cover}$.
Finally, the ratio $(A_{apx} - A_{cover})/A_{apx}$  represents the normalized ``room for improvement'' of the approximated solution. We call this measure the \emph{Optimality Gap}.

The optimality gap does not consider part of $A_{opt}$ that is not ``covered'' by $A_{apx}$, e..g, the orange shaded polygon $X'GDXX'$ in the example and which we denote as $A_{miss}$.  This, which we name \emph{Coverage Gap}, is a consequence of the skyline approximation, i.e., the more points it is missing, the larger such gap. In what follows we compute this (normalized) coverage gap as $(A_{miss}/A_{opt})$ and, like the optimality gap, the smaller it is, the better.

Finally, note that in \emph{all the experiments we conducted the optimality and coverage gaps were below 50\% and 15\% respectively}, quantifying that the approximated linear skylines were of good quality.
As well, the processing time of APX-PSD was always smaller than 1 second and around two orders of magnitude smaller than BSL-PSD's.

\subsubsection{Efficiency}

We evaluate the efficiency of our PSD-\NameProblem by comparing its processing time to the processing time of BSL\NameProblem. The experiments are designed to test whether APX-\NameProblem can provide a good solution in real-time. Even though we can not compare the BSL-\NameProblem and APX-\NameProblem in terms of effectiveness and efficiency for large store cardinality and shopping list size, we can confirm the scalability of APX-\NameProblem reporting the processing time for these extreme cases.

From Sections \ref{sec: baseline approach} and \ref{sec: approximated approach} recall that we assume some information -- namely the shortest path between stores and the stores partitioning -- is pre-computed offline. 
We argue that such assumption is reasonable, since stores are seldom added to or removed from networks. Moreover such pre-computation represents a not overly expensive one-time cost.

\hspace{1pt}
\begin{table}[htb]
  \begin{center}
    \begin{tabular}{l c c r} 
     \toprule
      \textbf{City} & \textbf{Shortest Path}& \textbf{Partitioning} & \textbf{Total}\\
       \midrule
      \vspace{3pt}
      Amsterdam & 31.40 & 1.56 & 32.96 \\
      \vspace{3pt}
      Oslo & 202.78 & 2.32 & 205.10\\
      \vspace{3pt}
      Berlin & 1072.31 & 4.76 & 1077.07 \\
      \bottomrule
    \end{tabular}
    \caption{{Pre-computation run-time (in seconds)}}
    \label{tab:Offdist}
  \end{center}
\end{table}

Table~\ref{tab:Offdist} shows the time required to pre-compute the shortest path between every pair of stores and also perform the store partitioning. Note that while we have used a typical implementation of Dijkstra's algorithm for shortest paths computation, more efficient alternatives could be used as this is a step completely independent of the approaches being proposed in our work.

\subsection{Effects of store cardinality}

\begin{figure} [htb]
   \centering
        \includegraphics[width=0.49\columnwidth]{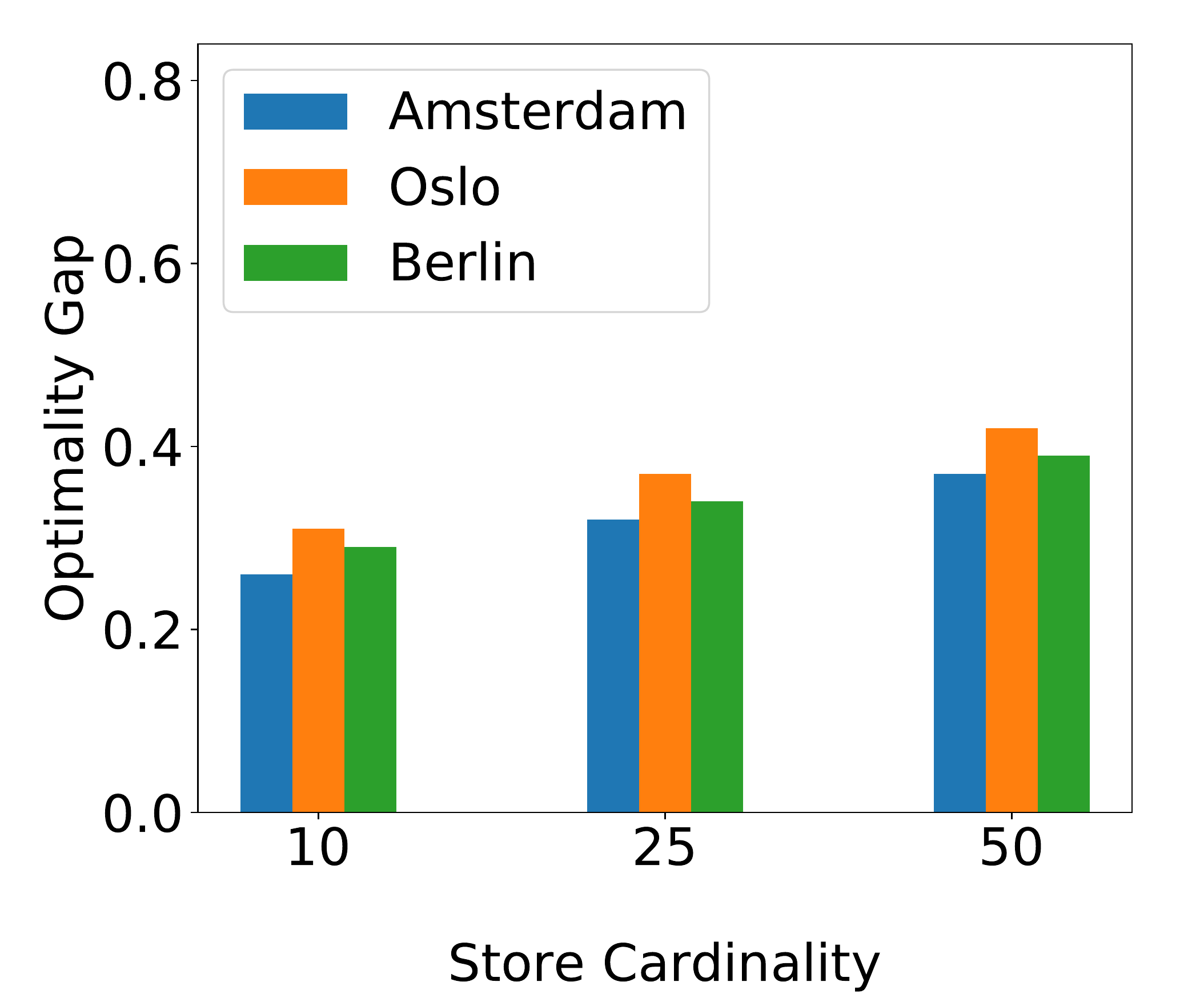}
        \includegraphics[width=0.49\columnwidth]{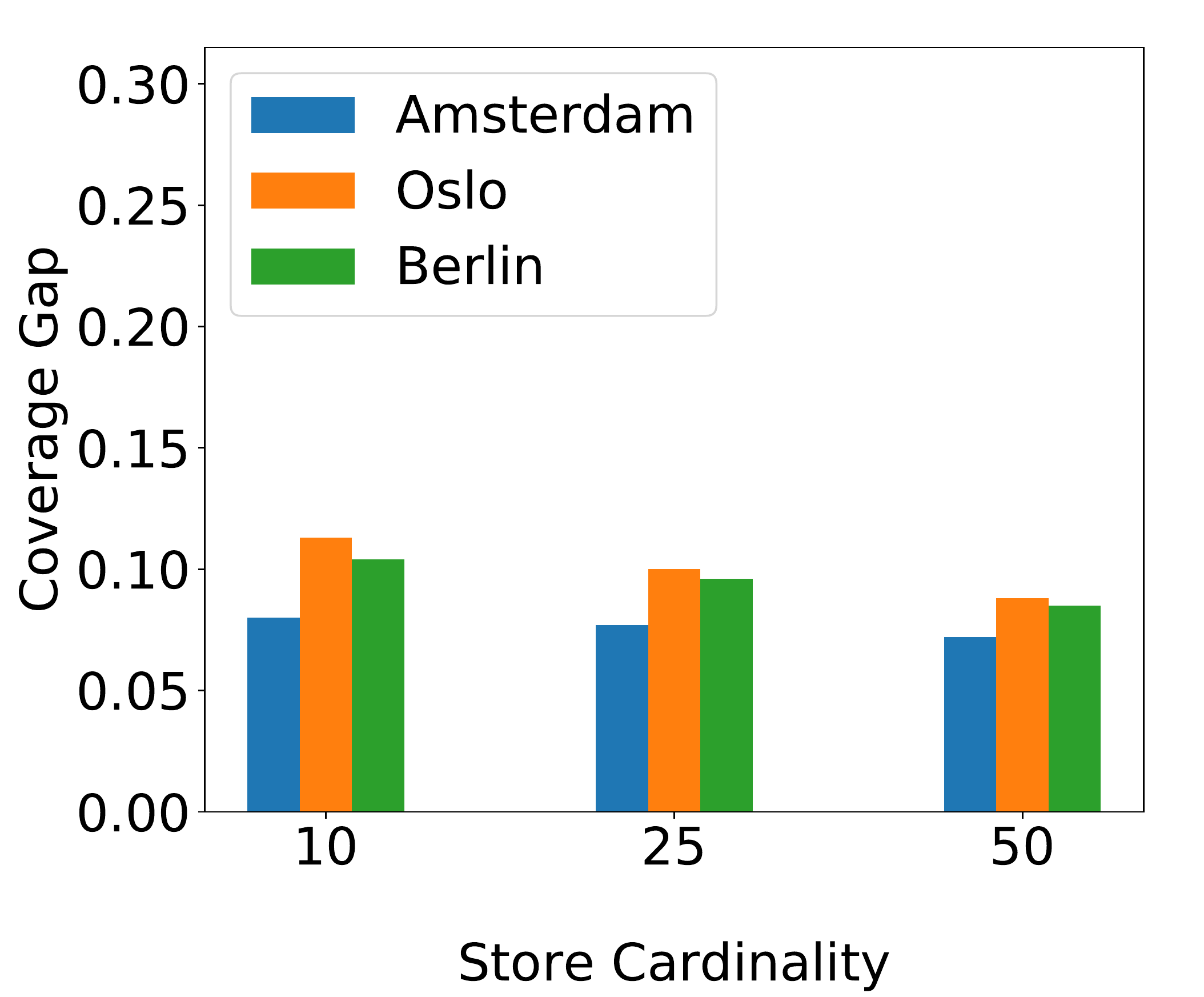}
    \caption{{Effectiveness w.r.t. store cardinality}}
   \label{fig:cc_card}
\end{figure}

Figure~\ref{fig:cc_card} shows that the optimality gap increases when increasing the store cardinality, while the coverage gap decreases.  
We explain such behaviour by observing how quad-trees as well as the scoring function react to changes in store cardinality. 
When store cardinality increases, the number of stores in each quad-tree leaf increases accordingly. 
Since product costs are distributed uniformly in the default case, adding more stores smooths the average cost of each product in different partitions, which in turn reduces the impact of product costs in APX-\NameProblem's scoring function. 
Therefore, the linear skyline will include routes with higher costs, which will decrease the coverage gap. Consequently, APX-\NameProblem has to visit more leaves to complete the shopping list. Furthermore, with larger store cardinality BSL-\NameProblem generates shopping routes with lower shopping time that APX-\NameProblem fails to find, thus increasing the optimality gap.

\begin{figure}[htb]
   \centering
        \includegraphics[width=0.49\columnwidth]{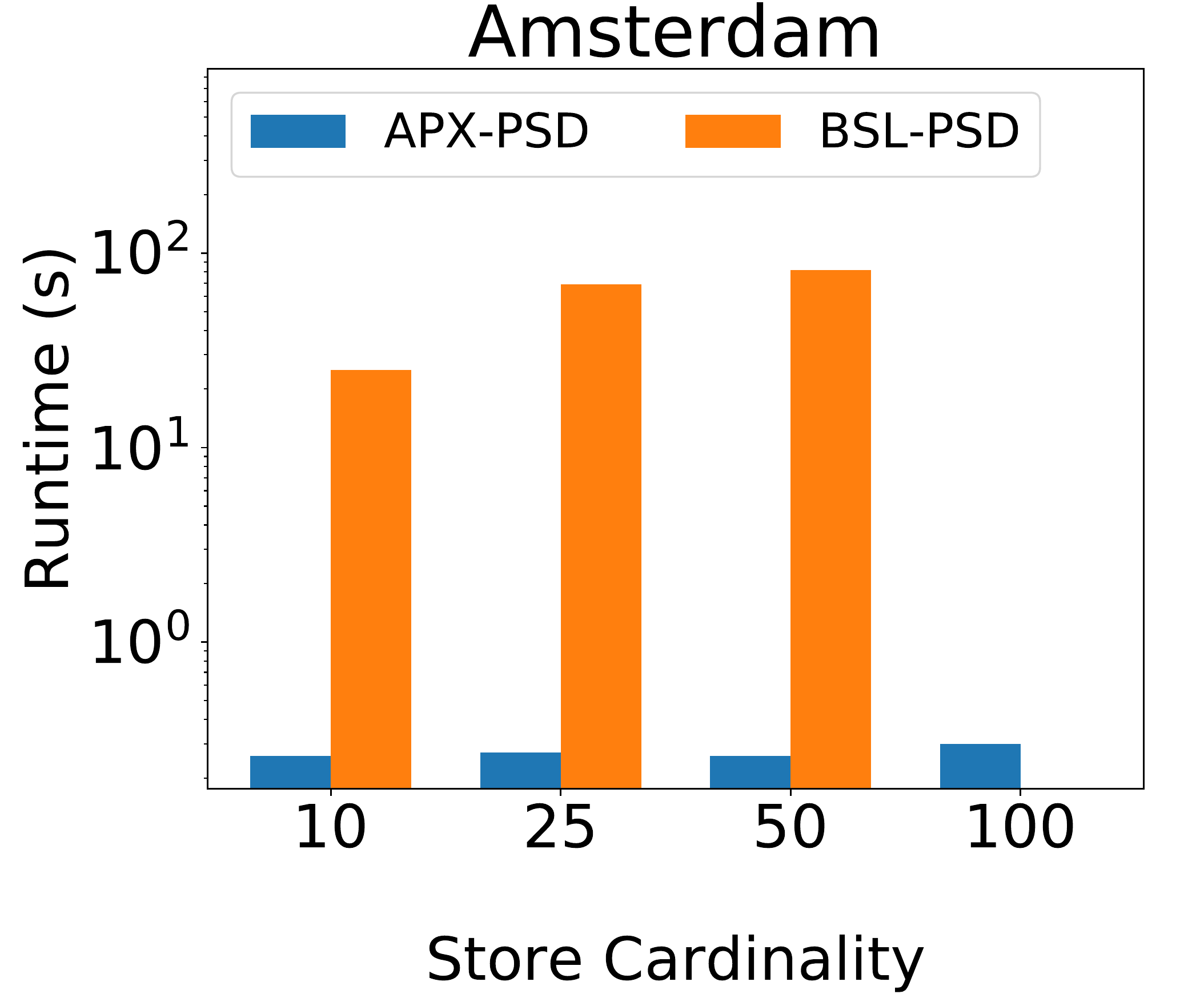}
        \includegraphics[width=0.49\columnwidth]{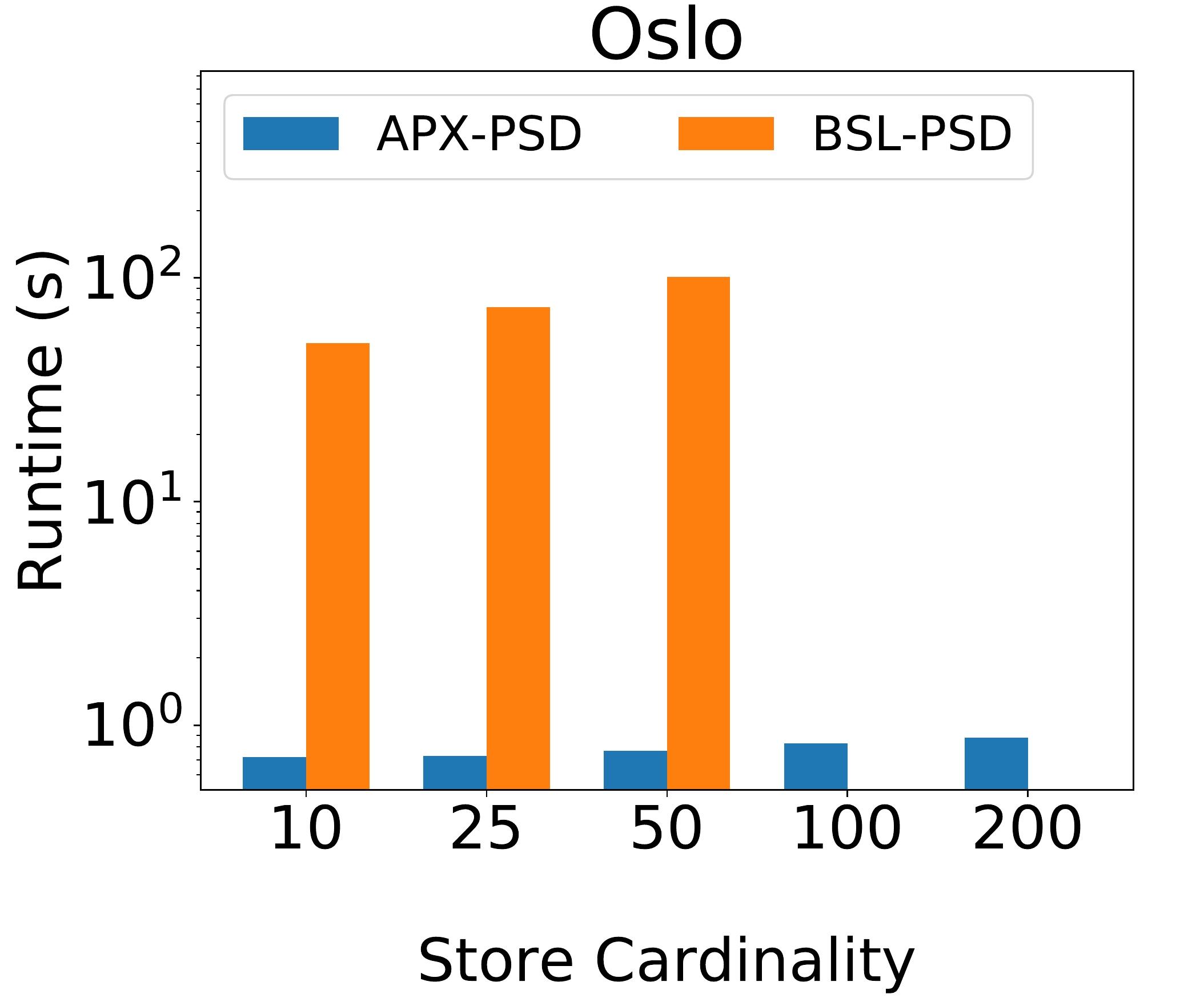}
        \includegraphics[width=0.49\columnwidth]{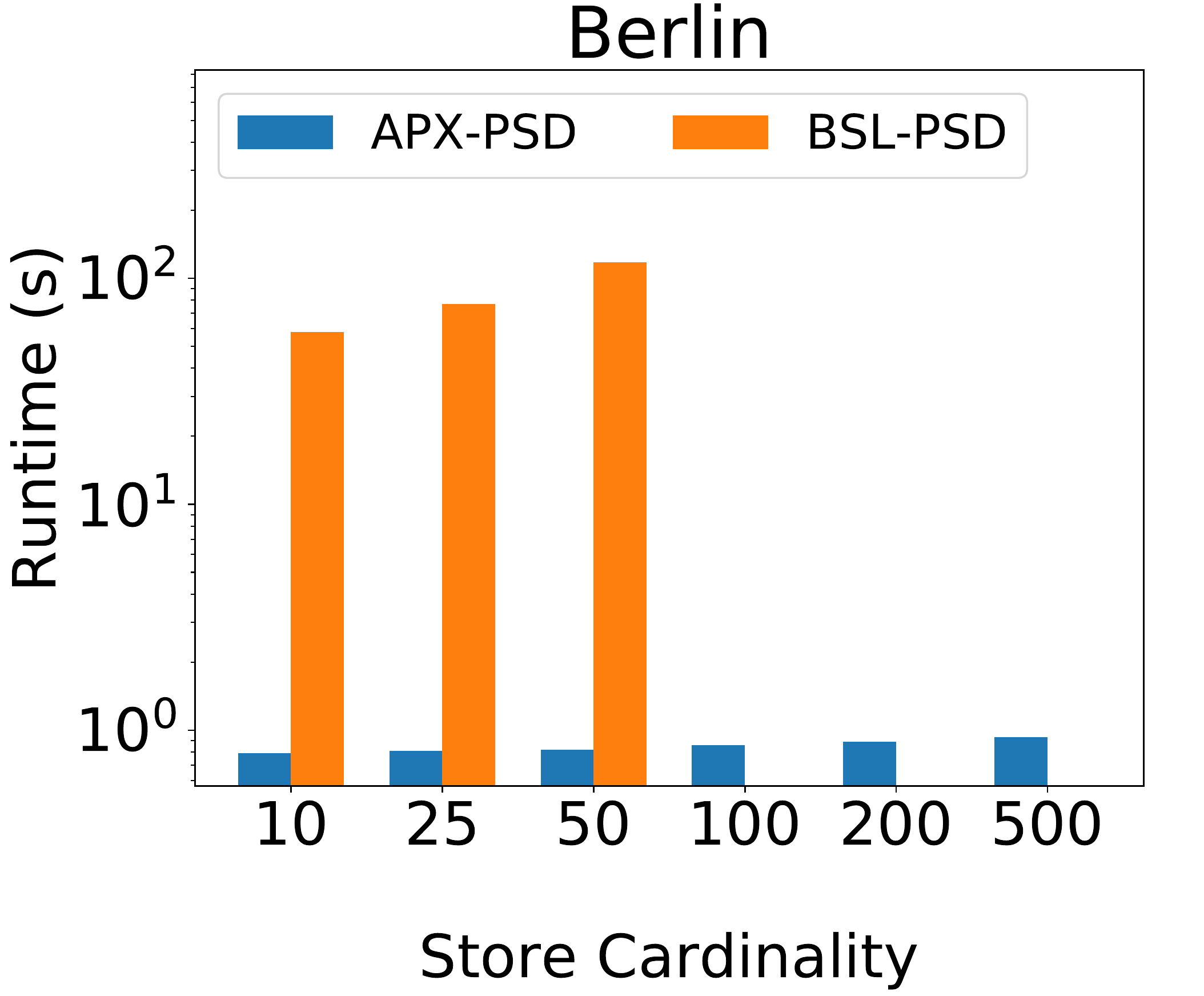}
   \caption{{Efficiency w.r.t. store cardinality.}}
   \label{fig:t_card}
\end{figure}

Figure~\ref{fig:t_card} shows that BSL-\NameProblem's processing time increases when store cardinality increases, mainly due to an increased number of stores (and thus potential candidates) to consider.
On the other hand, APX-\NameProblem's processing time exhibits small changes. Recall that the leaves' capacity in a quad-tree is fixed, thus increasing the store cardinality increases the tree's depth.
Notice that, we varied the parameter upto $500$ stores for Berlin but only up to $200$ stores and $100$ stores for Oslo and Amsterdam respectively. We could not test Amsterdam or Oslo with larger store cardinality due to the limitation of the datasets respectively, 
Recall that APX-\NameProblem's time-dominant component deals with shopping routes generation and evaluation, rather than tree traversal, which explains the small impact on performance. 

\subsection{Effects of product cost distribution}

\begin{figure} [htb]
   \centering
        \includegraphics[width=0.49\columnwidth ]{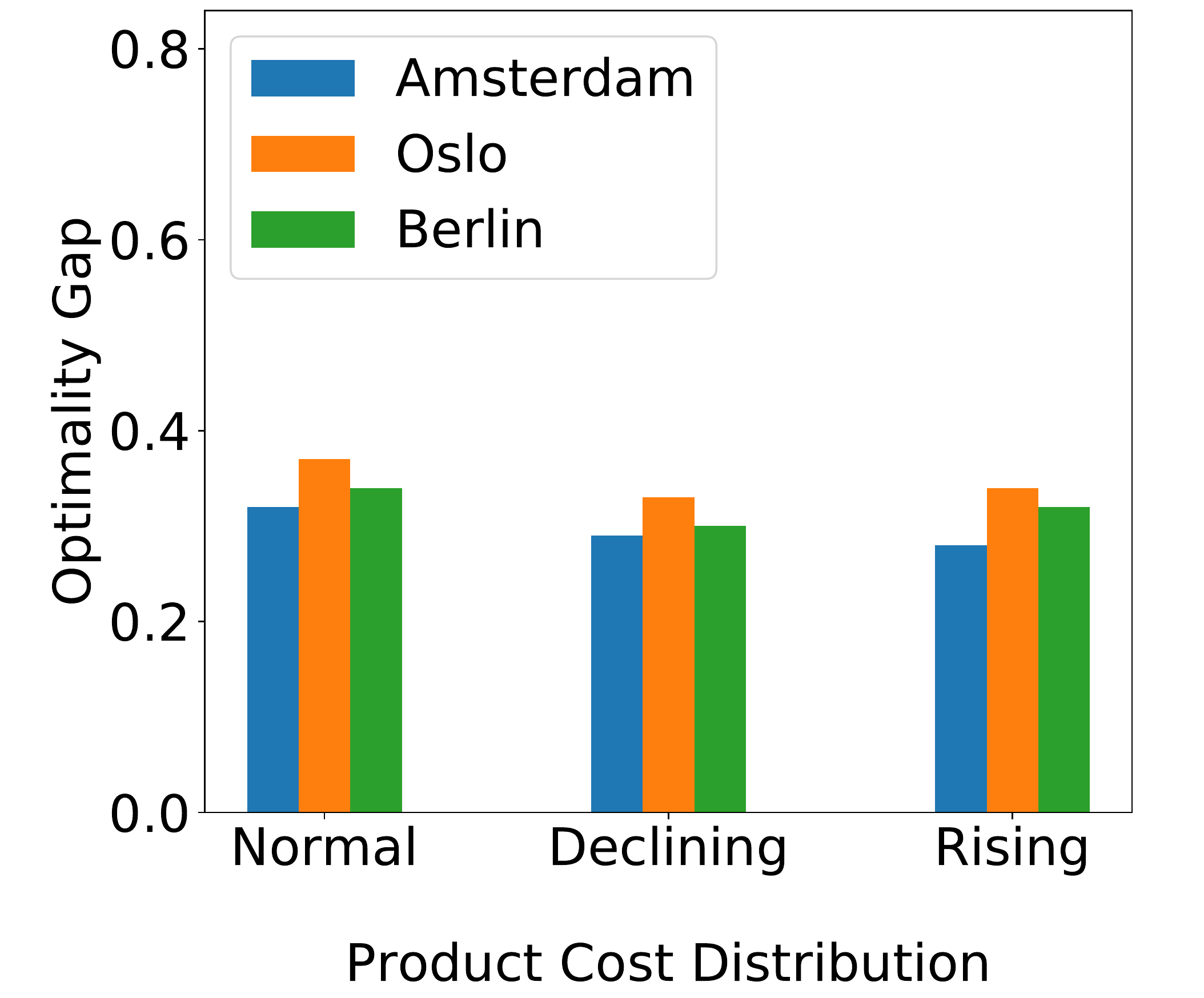}
        \includegraphics[width=0.49\columnwidth]{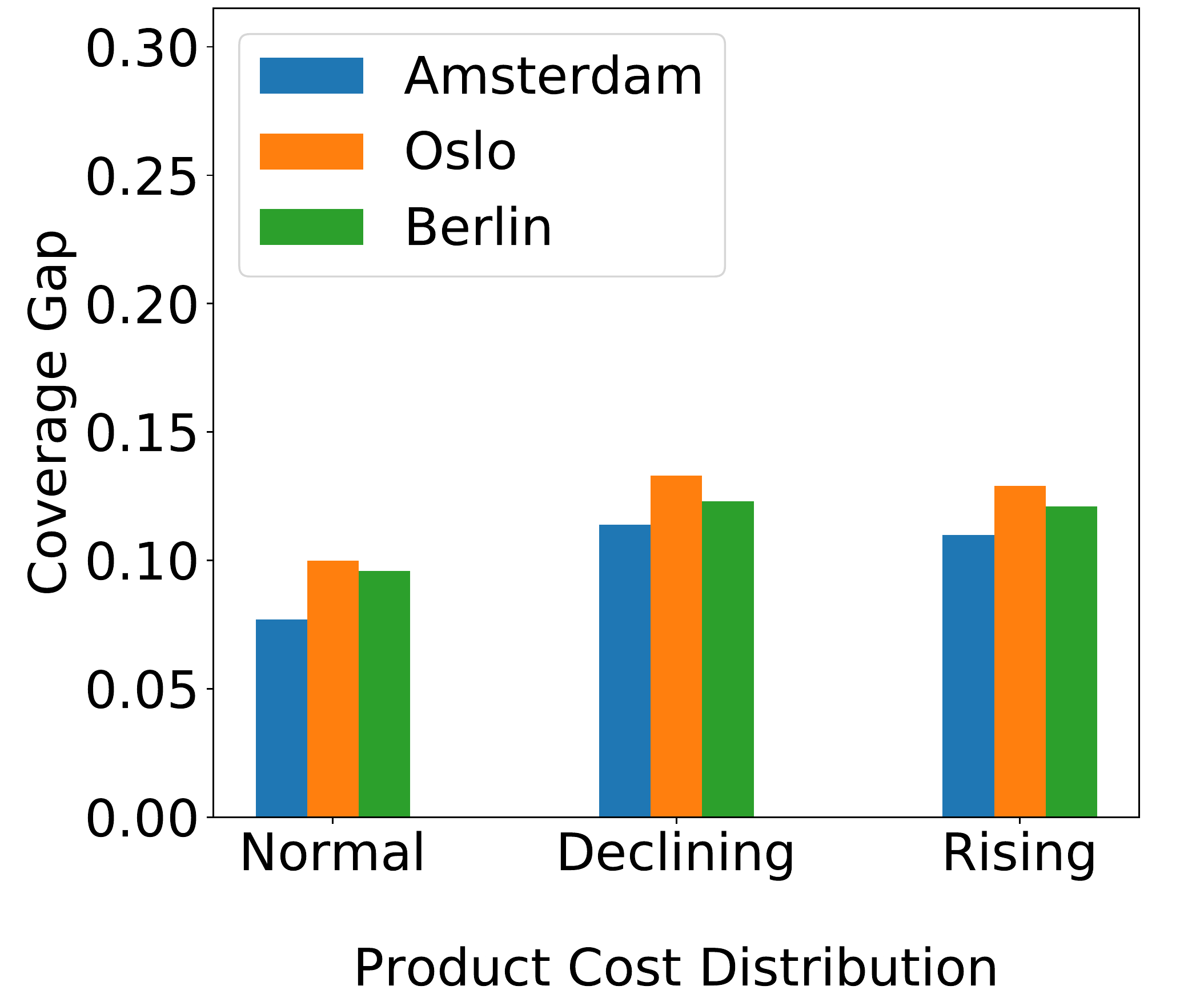}\\
   \caption{{Effectiveness w.r.t. different cost distributions.}}
   \label{fig:cc_cost}
\end{figure}

In Figure~\ref{fig:cc_cost} we can see that the ``Declining'' and ``Rising'' cases have comparable optimality gaps and higher coverage gaps to the ``Normal'' one. We argue that these results are due to the characteristics of the scoring function used by APX-\NameProblem which, we recall, attempts to balance the importance given to shopping time and shopping cost. 
Since the product cost gradually decreases towards one direction for both ``Declining" and ``Rising", the scoring function manages to minimize cost better than ``Normal" distribution. However, the optimality gap created due to creating partial routes from a leaf at a time remains comparable to the ``Normal" distribution.

Interestingly, both the optimality gap and coverage gap are the highest for Oslo in all of the cases, which we attribute to the skewed distribution of stores compared to the other cities (Figure~\ref{fig:maps}). 

\begin{figure} [htb]
   \centering
        \includegraphics[width=0.49\columnwidth]{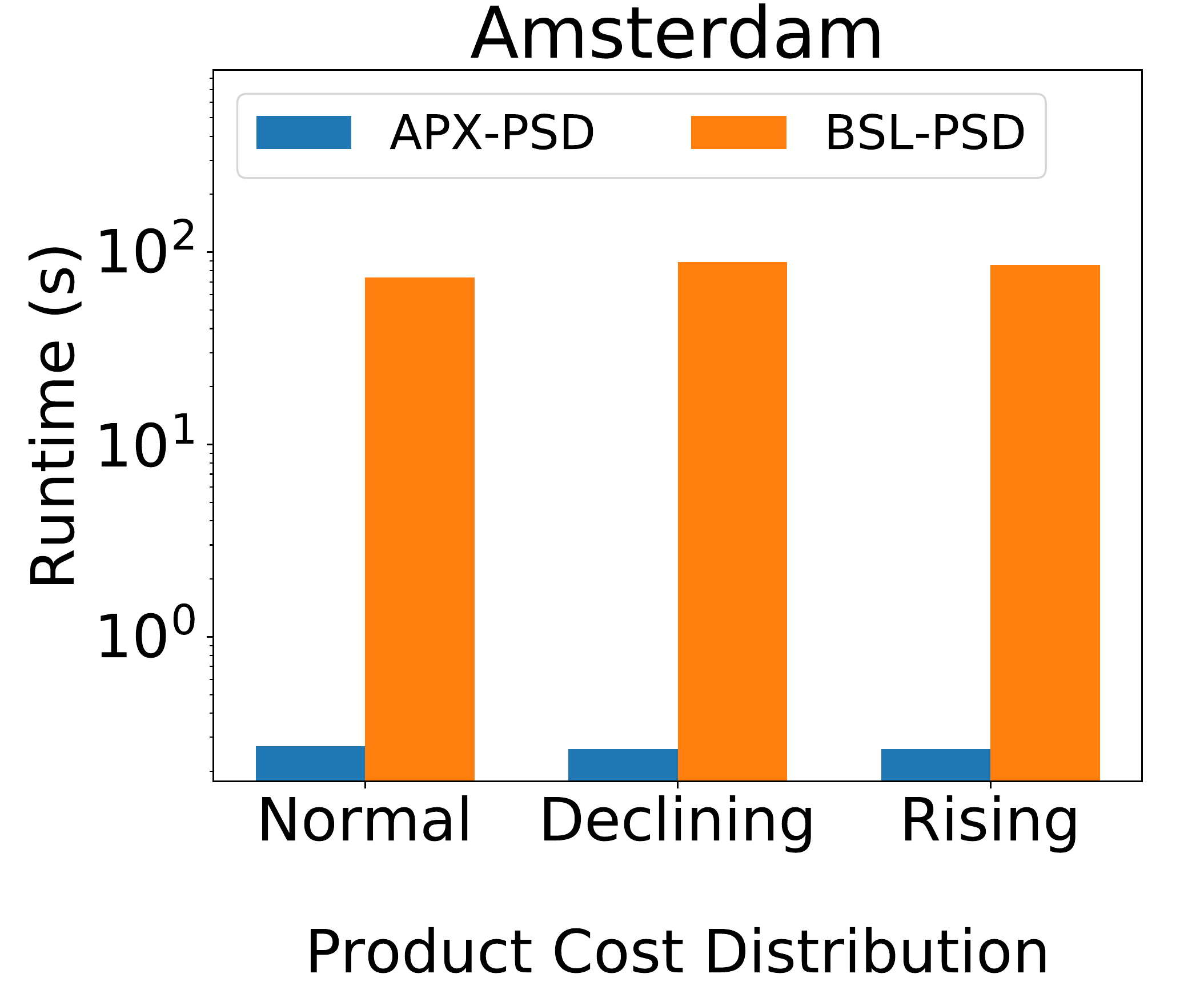}
        \includegraphics[width=0.49\columnwidth]{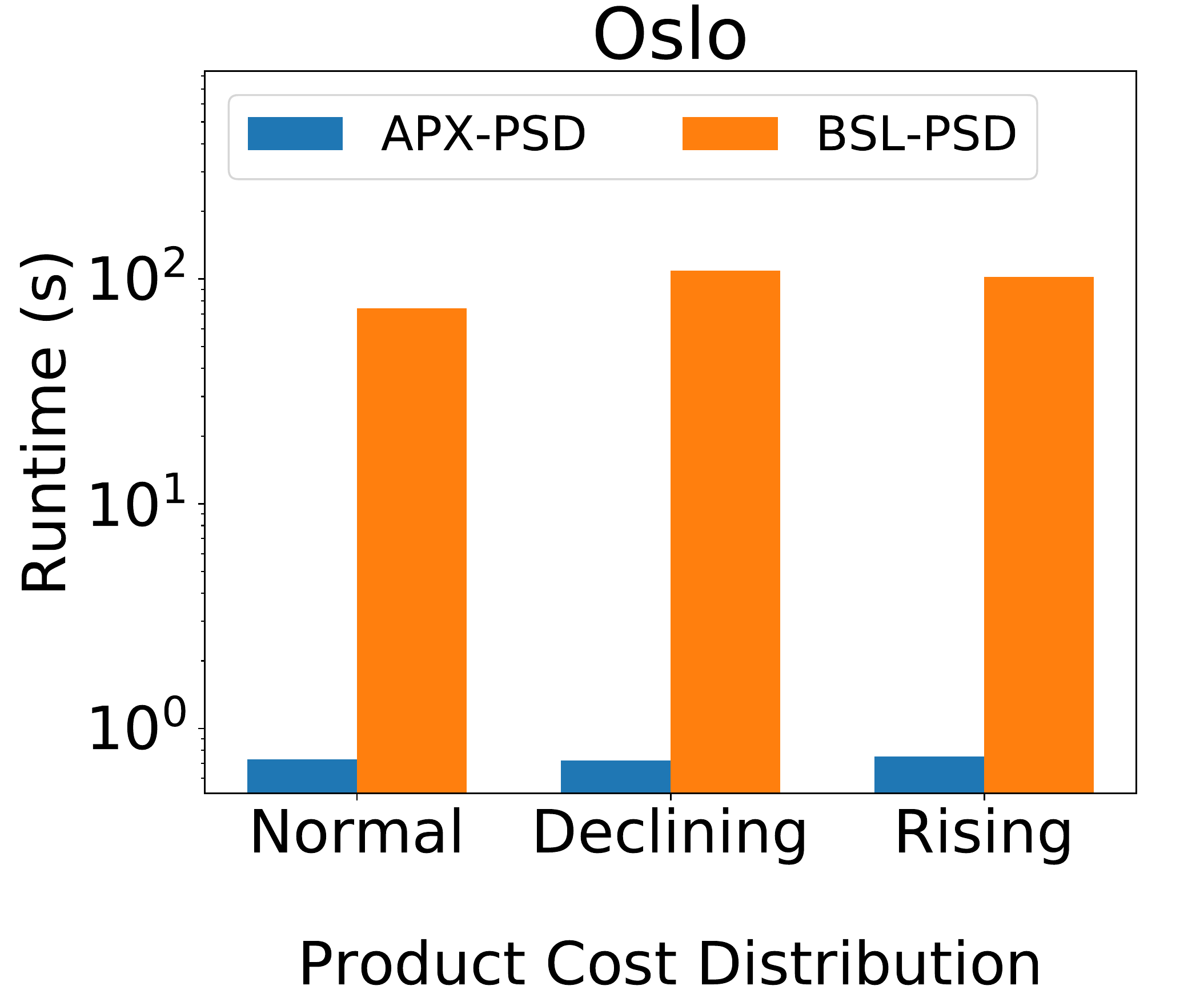}
        \includegraphics[width=0.49\columnwidth]{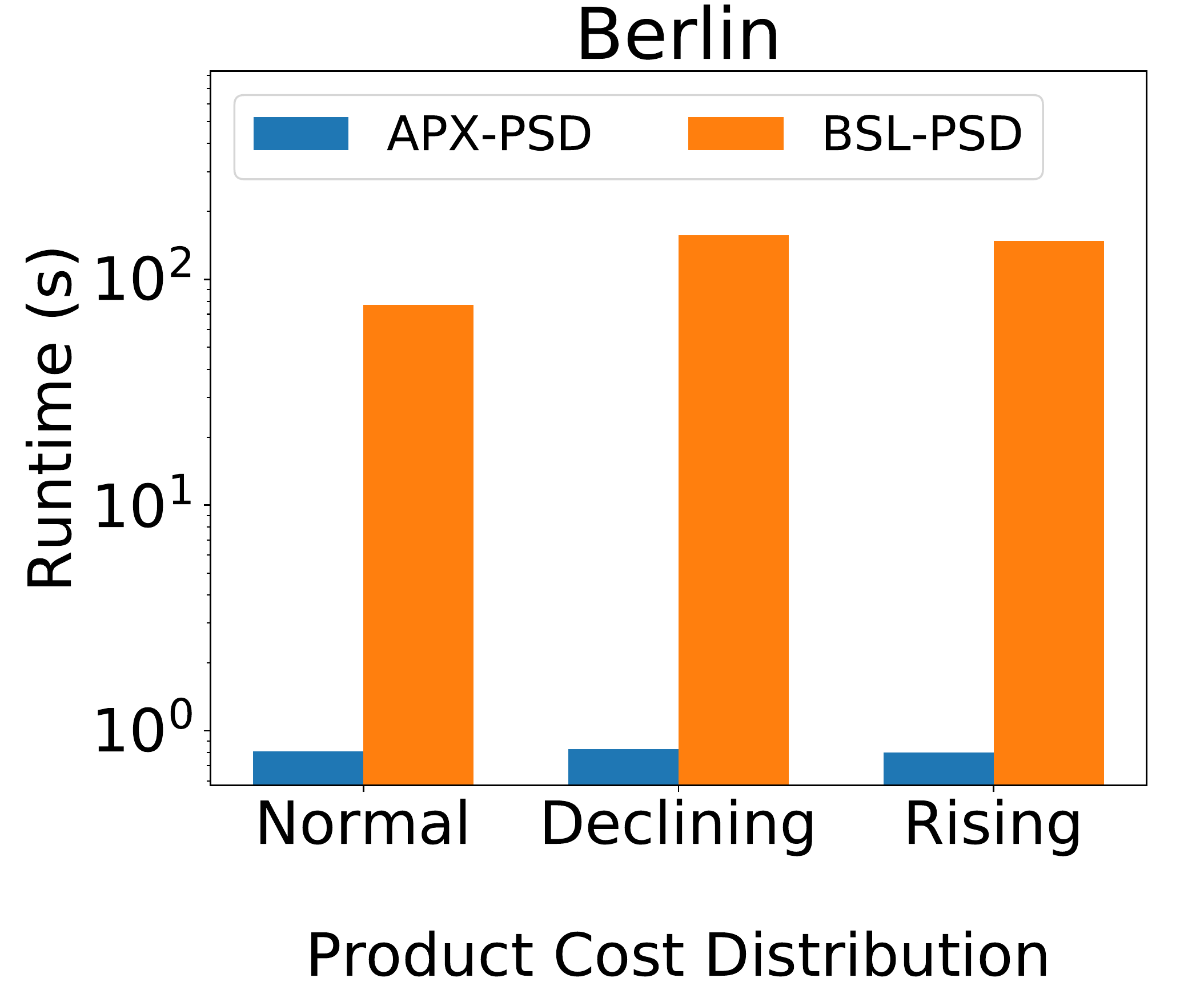}
   \caption{{Efficiency w.r.t. different cost distributions.}}
   \label{fig:t_cost}
\end{figure}

Figure \ref{fig:t_cost} shows that BSL-\NameProblem's processing time using Berlin's network is larger for the ``Declining'' and ``Rising'' cases than for the ``Uniform'' one. 
In those two cases the low-cost products are distributed in certain areas of the map, and thus it is more likely to take longer to find routes with lower shopping costs if the shopper's and customer's delivery locations are far from those regions. Note that such differences are further amplified by Berlin's large network size.
However, APX-\NameProblem's processing time is not affected since it takes great advantage of pre-computed aggregated paths as it traverses the quad-tree hosting the store partitions.


\subsection{Effects of store size spatial distribution}

\begin{figure}
   \centering
        \includegraphics[width=0.49\columnwidth]{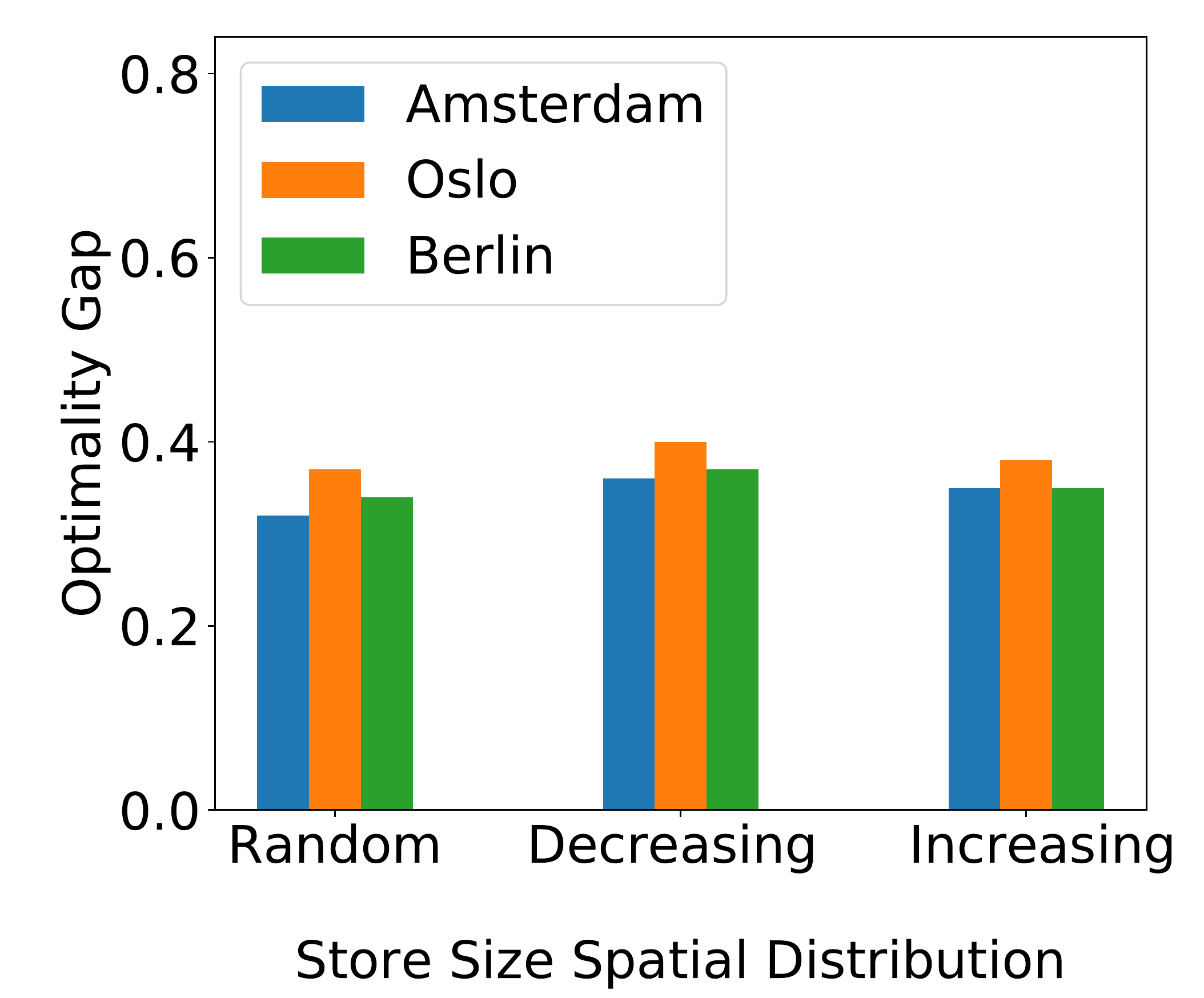}
        \includegraphics[width=0.49\columnwidth]{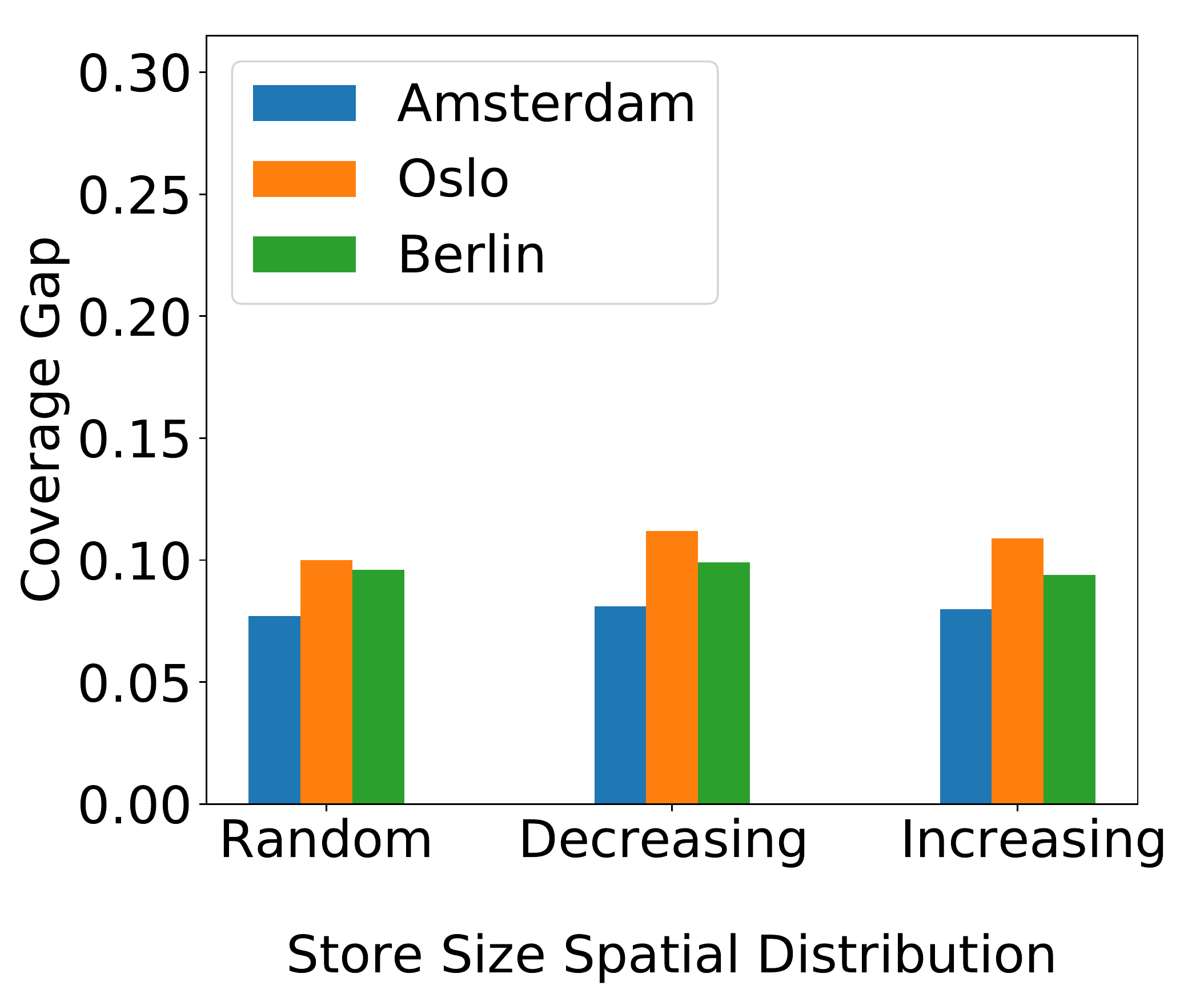}
   \caption{{Effectiveness w.r.t. different distributions of store size.}}
   \label{fig:cc_ss}
\end{figure}

In Figure~\ref{fig:cc_ss} we can see that the ``Decreasing'' and ``Increasing'' cases have comparable coverage and optimality gaps to the ``Random'' one. These results can be explained by taking into account the characteristics of APX-\NameProblem's scoring function, which makes it insensitive to different distributions of store sizes (apart from the effects that can be observed on the shape of both optimal and approximated skylines).

\begin{figure}
   \centering
        \includegraphics[width=0.49\columnwidth]{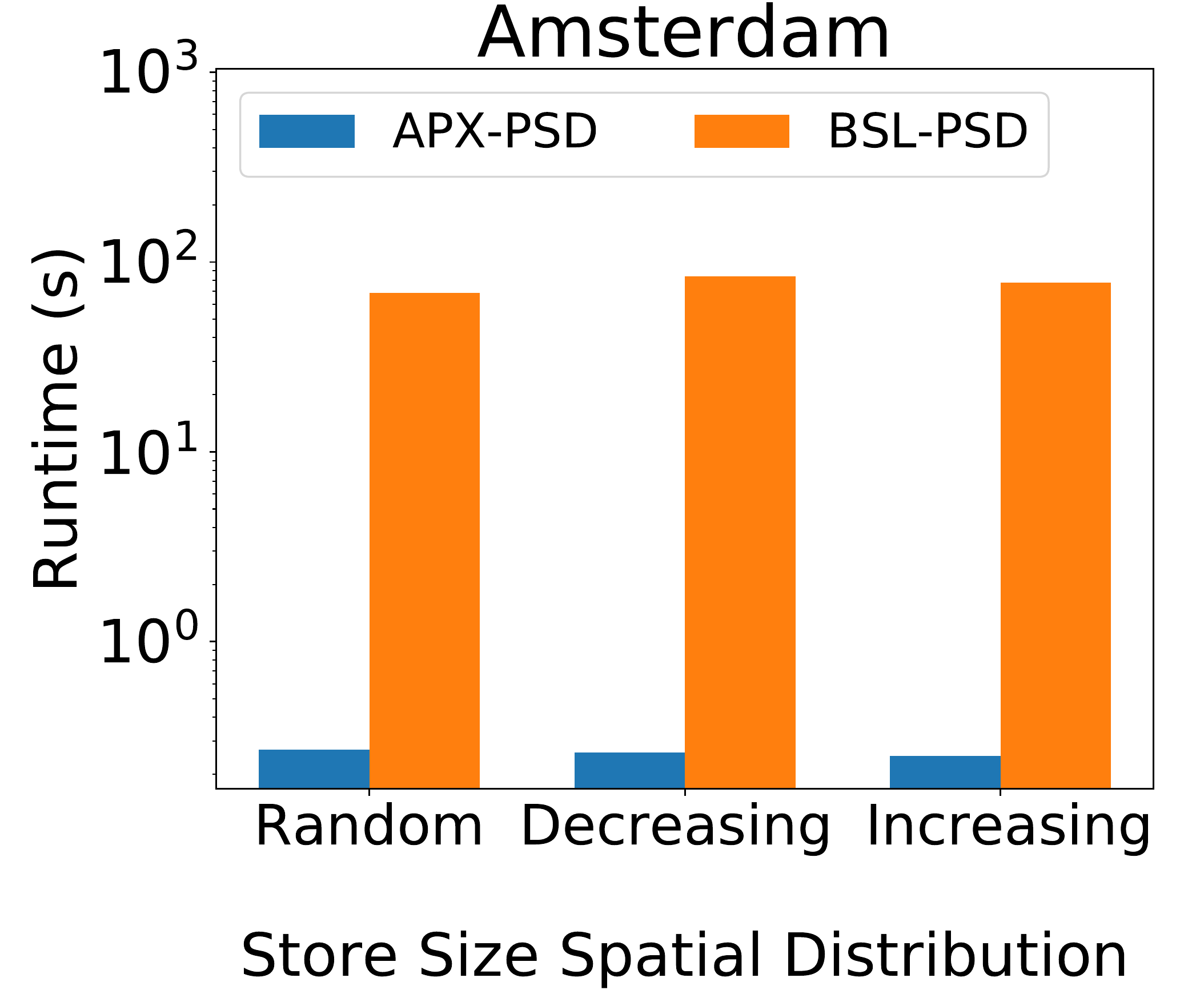}
        \includegraphics[width=0.49\columnwidth]{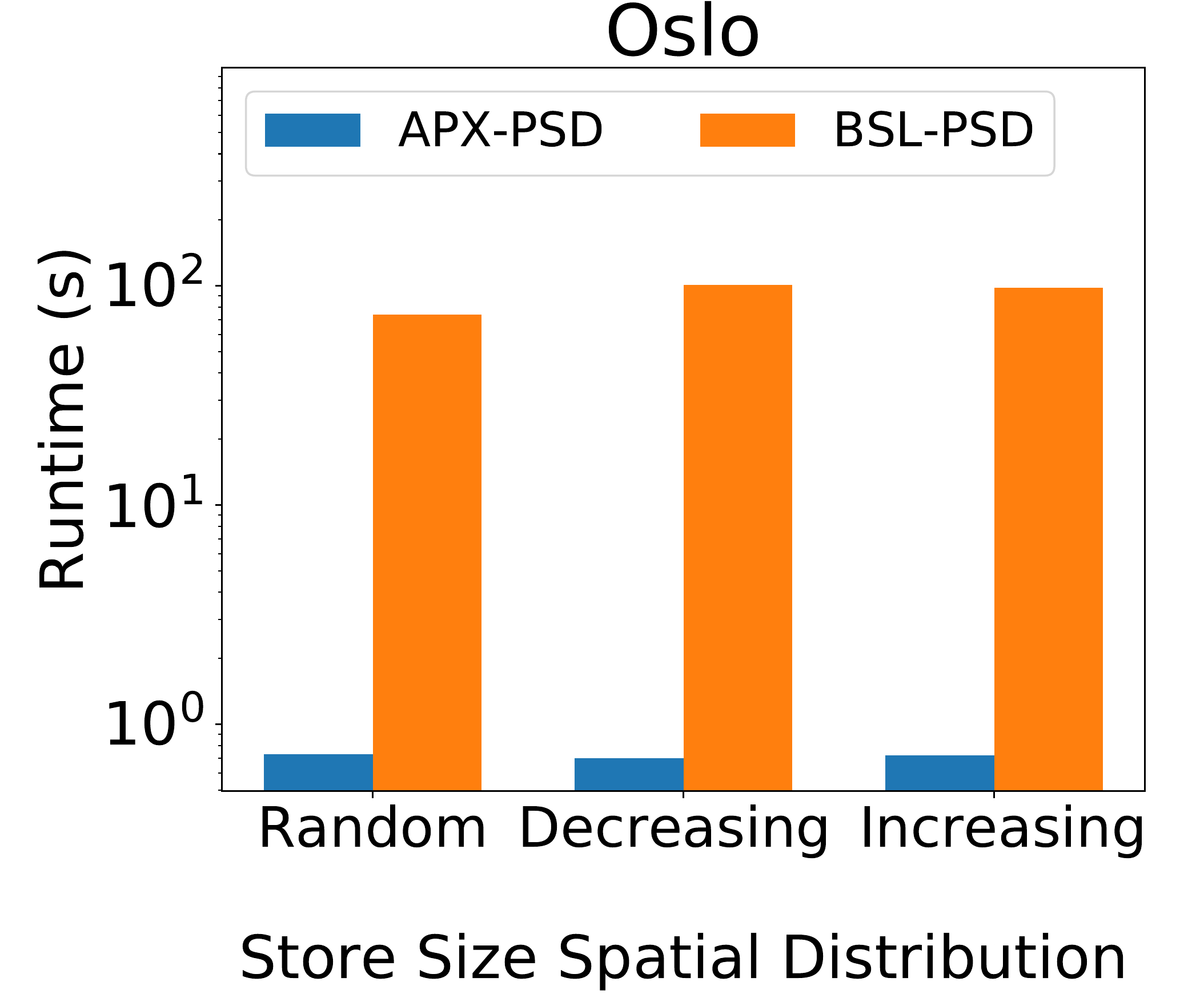}
        \includegraphics[width=0.49\columnwidth]{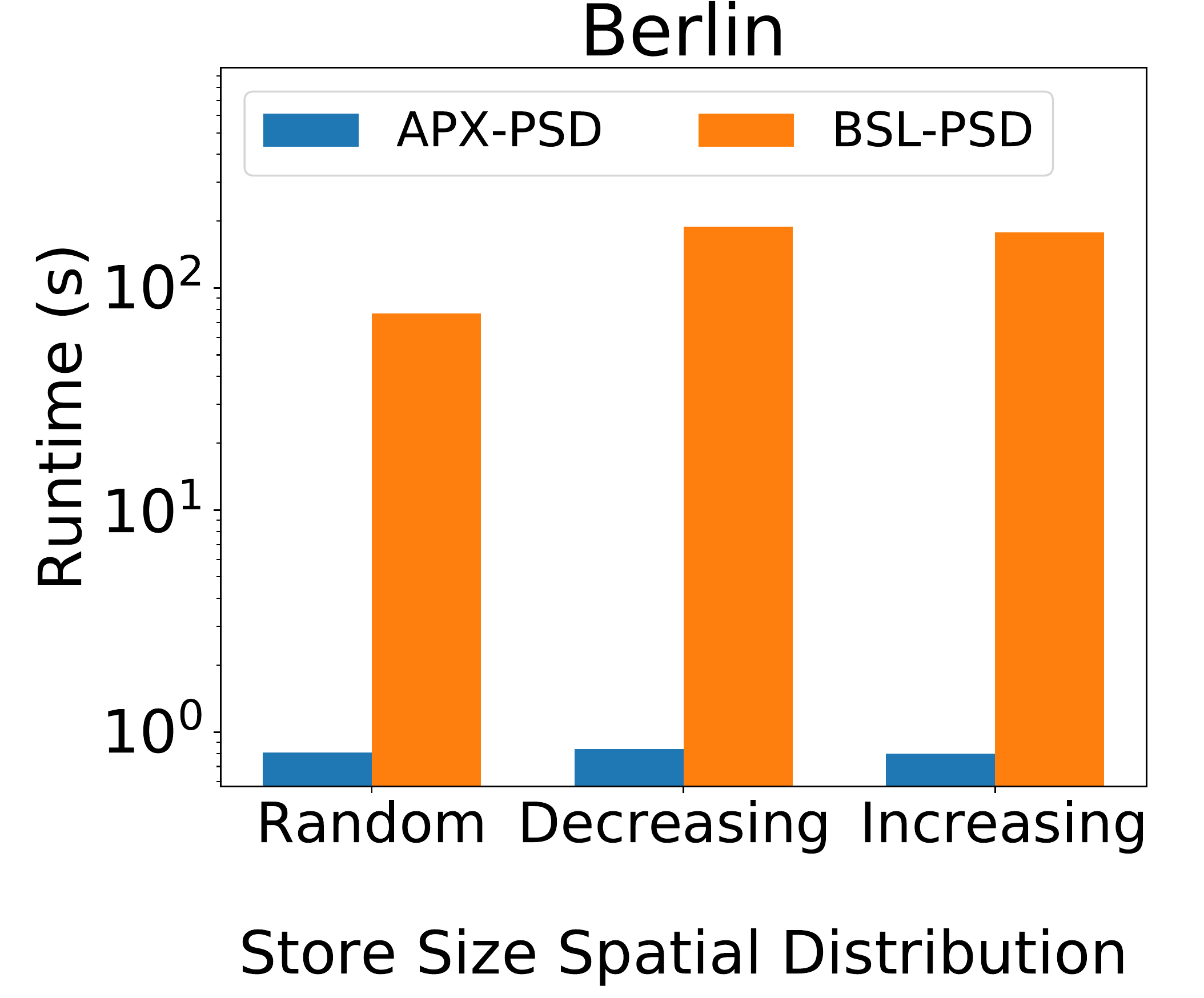}
   \caption{{Efficiency w.r.t. different distributions of store size.}}
   \label{fig:t_ss}
\end{figure}

BSL-\NameProblem's processing time (Figure \ref{fig:t_ss}) is higher when dealing with the ``Decreasing'' and``Increasing'' cases. Since the locations of larger stores are concentrated in certain areas, BSL-\NameProblem takes longer to generate routes with minimum cost and terminate (depending on the shopper's and customer's delivery locations), an effect that is further compounded by the city's network size. As usual, Berlin exhibits the largest processing time for the same reason observed with varying cost distributions. Finally note that APX-PSD's processing time remains unaffected.

\subsection{Effects of shopping list size}

\begin{figure} [htb]
   \centering
        \includegraphics[width=0.49\columnwidth]{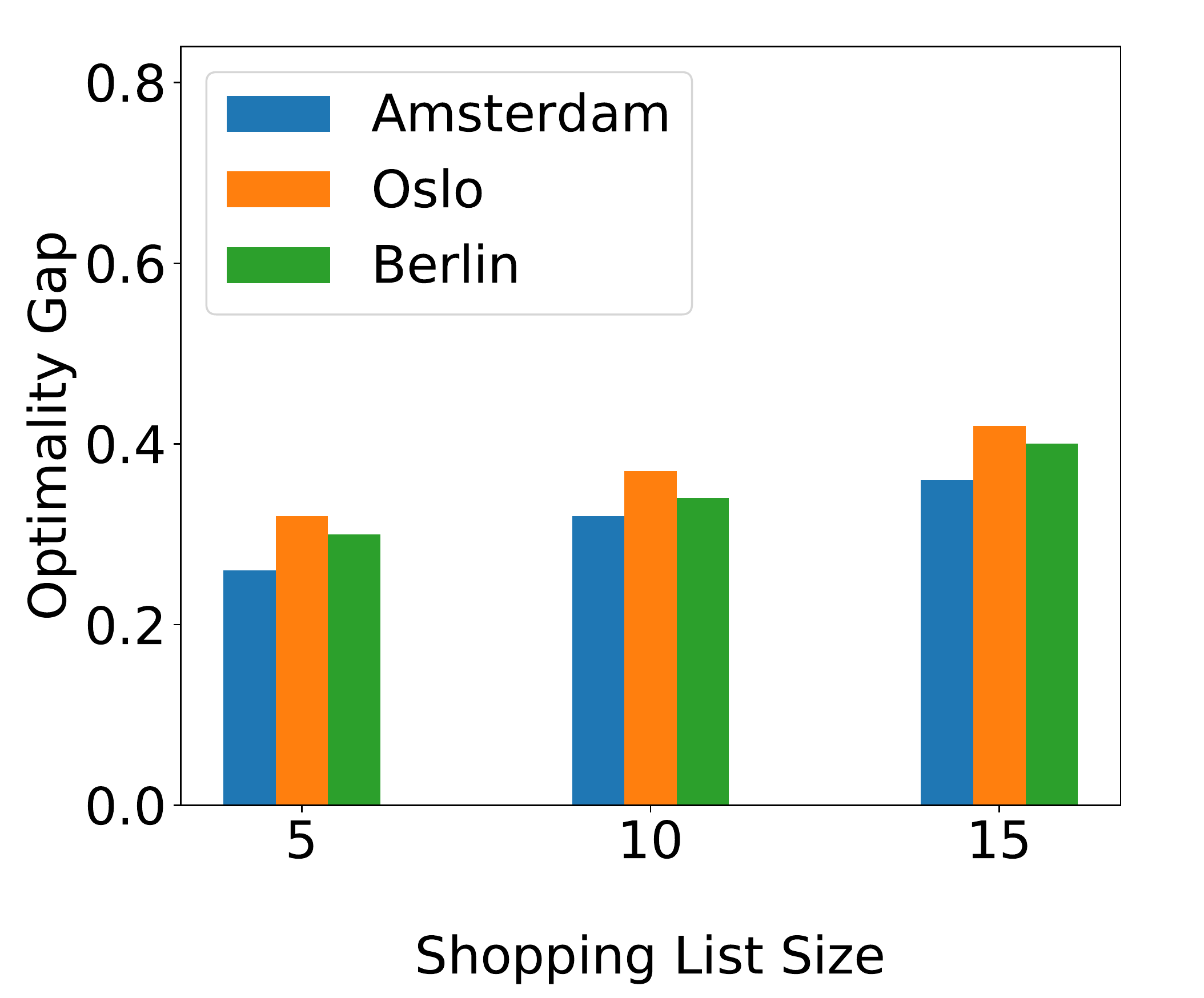}
        \includegraphics[width=0.49\columnwidth]{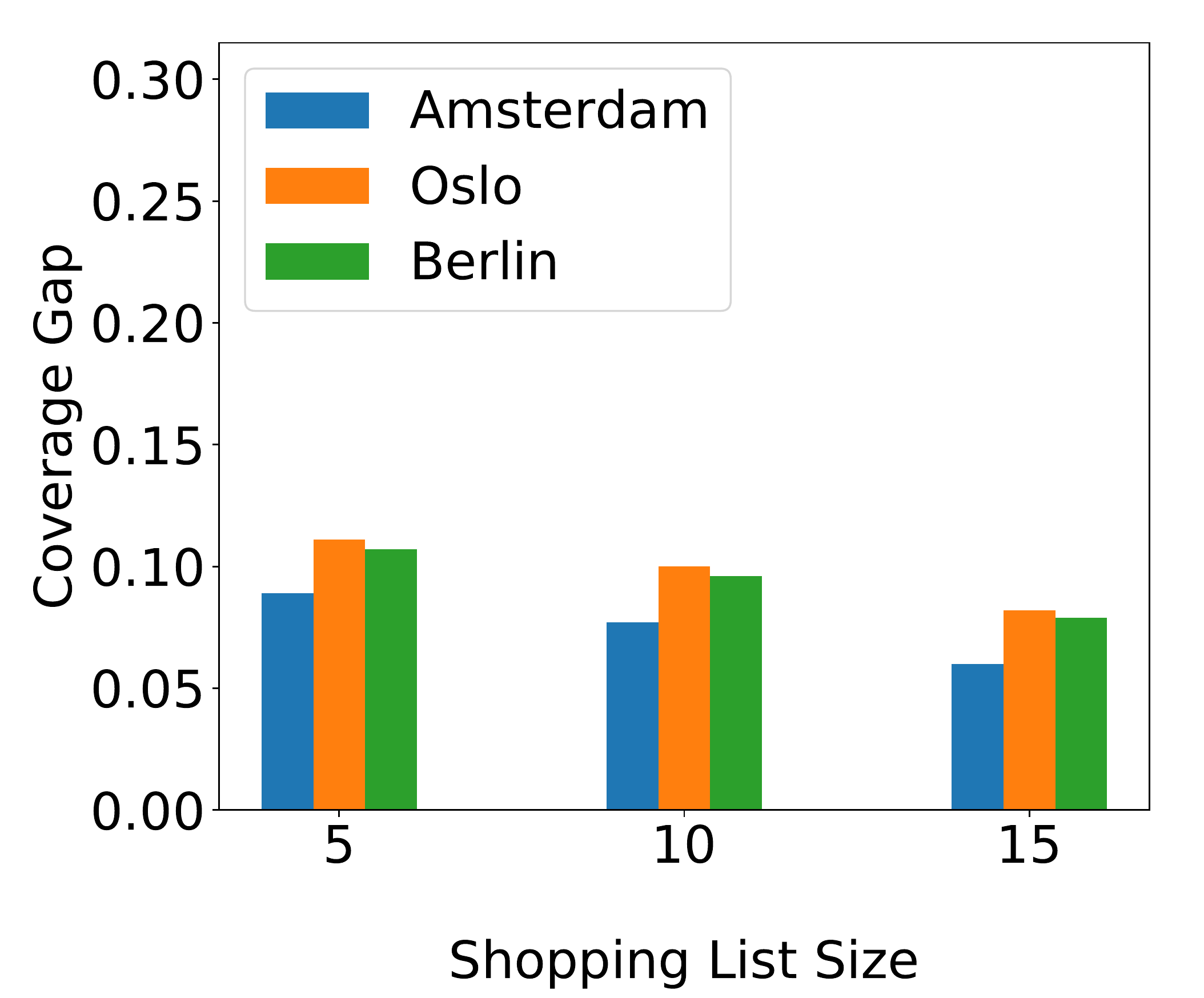}
   \caption{{Effectiveness w.r.t. shopping list size.}}
   \label{fig:cc_sl}
\end{figure}

The optimality (coverage) gap increases (decreases) with the shopping list size as evidenced in Figure \ref{fig:cc_sl}.  Larger shopping lists require more traversals of the network.  Recall that by construction, shopping routes are appended to existing partial routes.  Such appending means that previous not-so-good choices remain and their effect are further compounded by new potentially not-so-good choices as the algorithm evolves, worsening the approximation.
As a result both the shopping time and cost increase which increases the optimality gap and decreases the coverage gap.

\begin{figure} [htb]
   \centering
        \includegraphics[width=0.49\columnwidth]{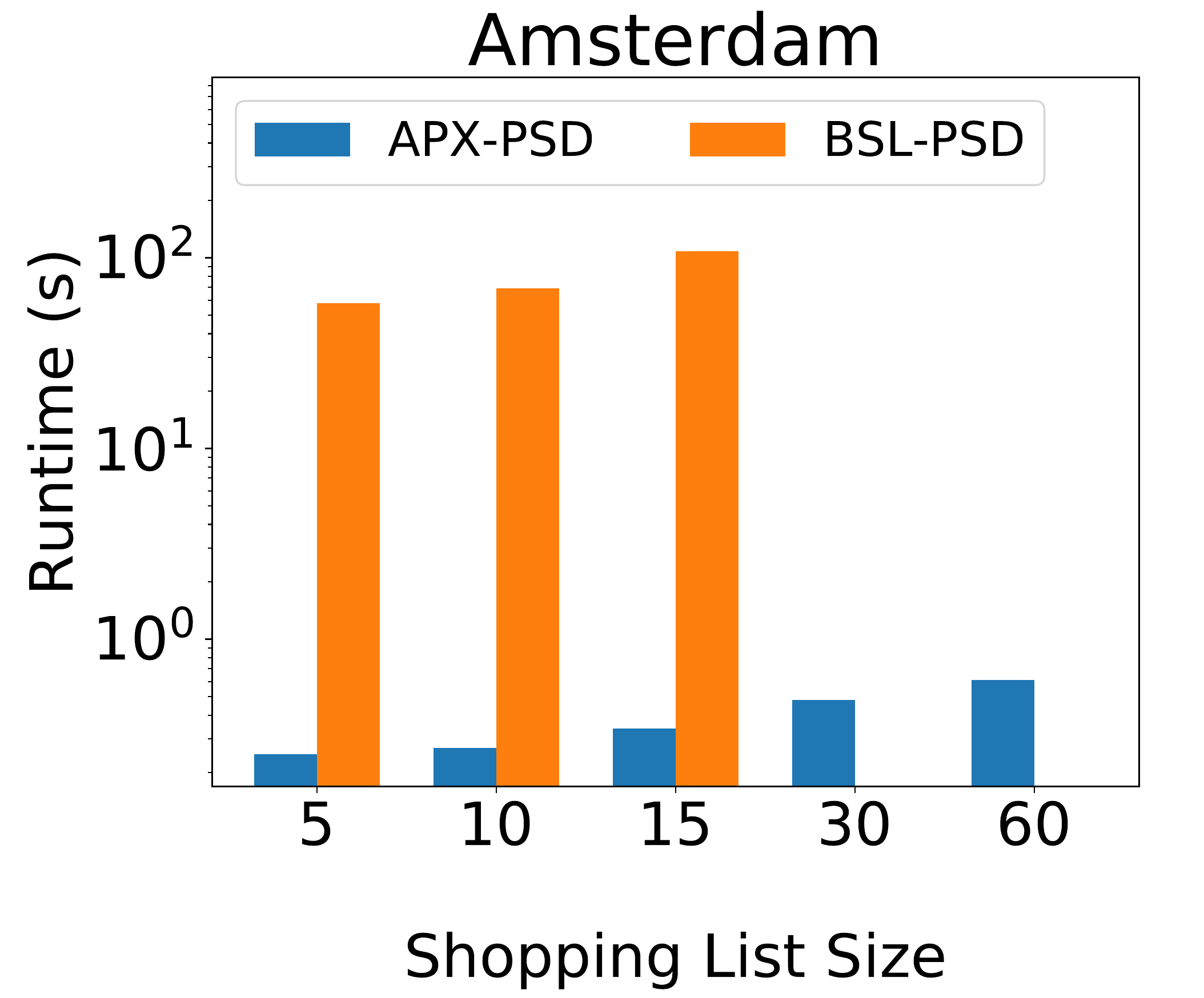}
        \includegraphics[width=0.49\columnwidth]{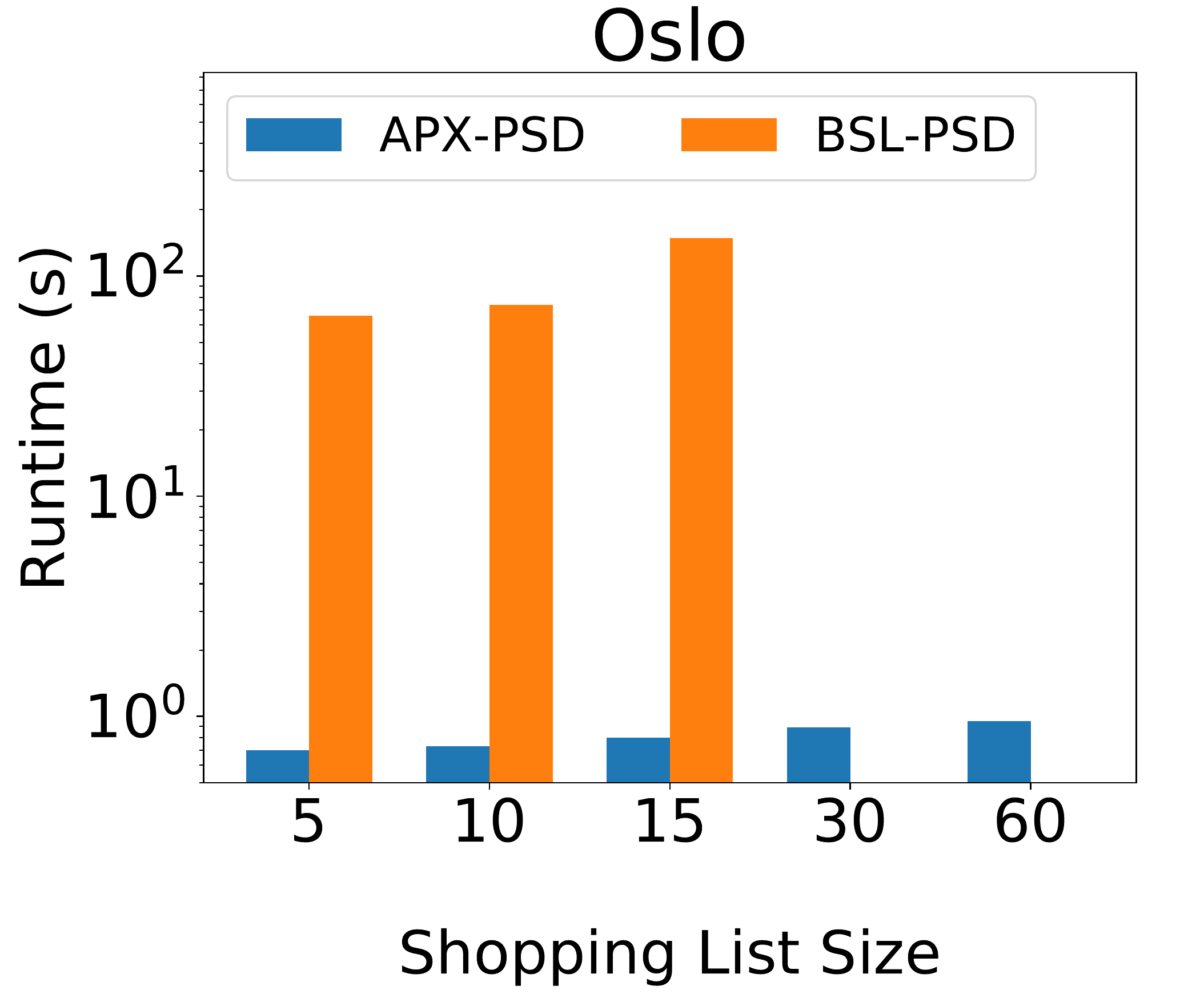}
        \includegraphics[width=0.49\columnwidth]{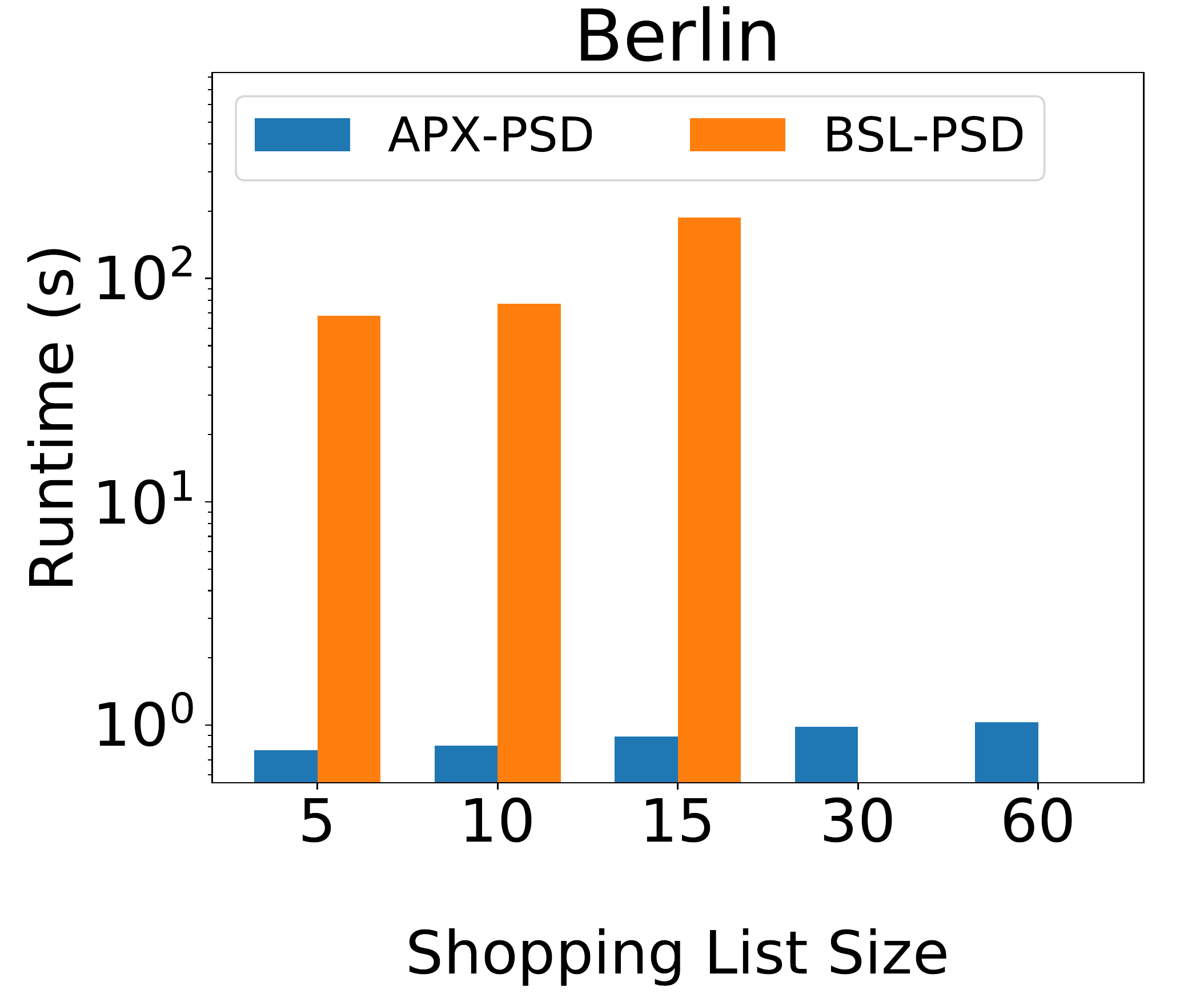}
   \caption{{Efficiency w.r.t. shopping list size.}}
   \label{fig:t_sl}
\end{figure}

BSL-\NameProblem's processing time (Figure \ref{fig:t_sl}) increases when increasing the shopping list size. This can be explained by observing that, on average, large shopping lists require more stores per route to be satisfied, and thus likely require to evaluate more candidate routes. 
On the other hand APX-\NameProblem is mildly affected by such increase, thanks to the noticeably smaller number of candidates it generates and evaluates by design.

\subsection{Effects of quad-tree leaf capacity}

\begin{figure}
   \centering
        \includegraphics[width=0.490\columnwidth]{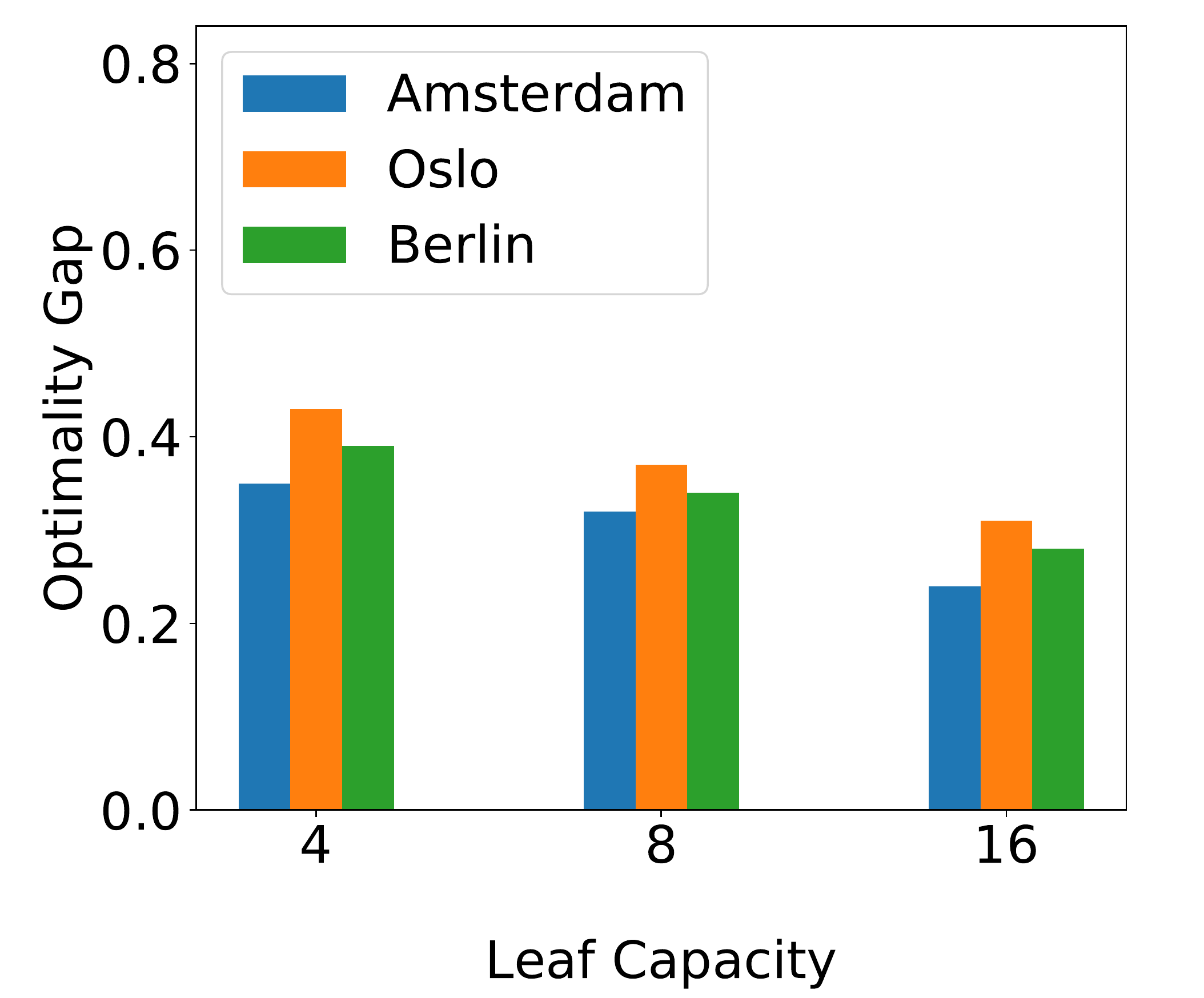}
        \includegraphics[width=0.490\columnwidth]{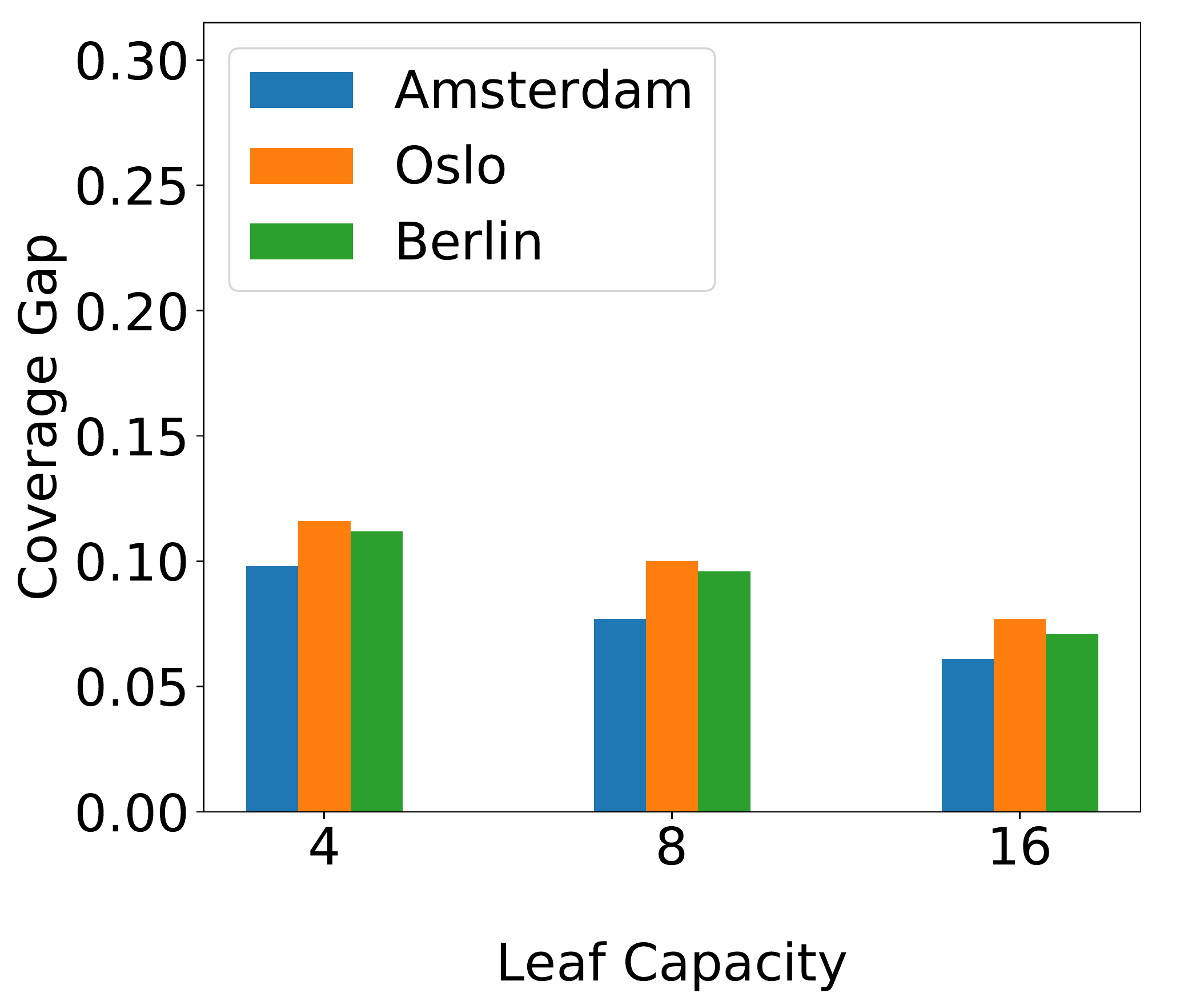}

   \caption{{Effectiveness w.r.t. varying quad-tree leaf capacity }}
   \label{fig:cc_leaf}
\end{figure}

From the results shown in Figure \ref{fig:cc_leaf} we see that the optimality and coverage gaps decrease (i.e., the overall result quality increases) as the leaf capacity increases. This can be explained by observing that leaves containing more stores allow APX-\NameProblem to generate and evaluate more partial shopping routes, thus allowing to discover shopping routes with increasingly lower costs (and thus closer to the ones computed by the baseline). 

\begin{figure}
   \centering
        \includegraphics[width=0.49\columnwidth]{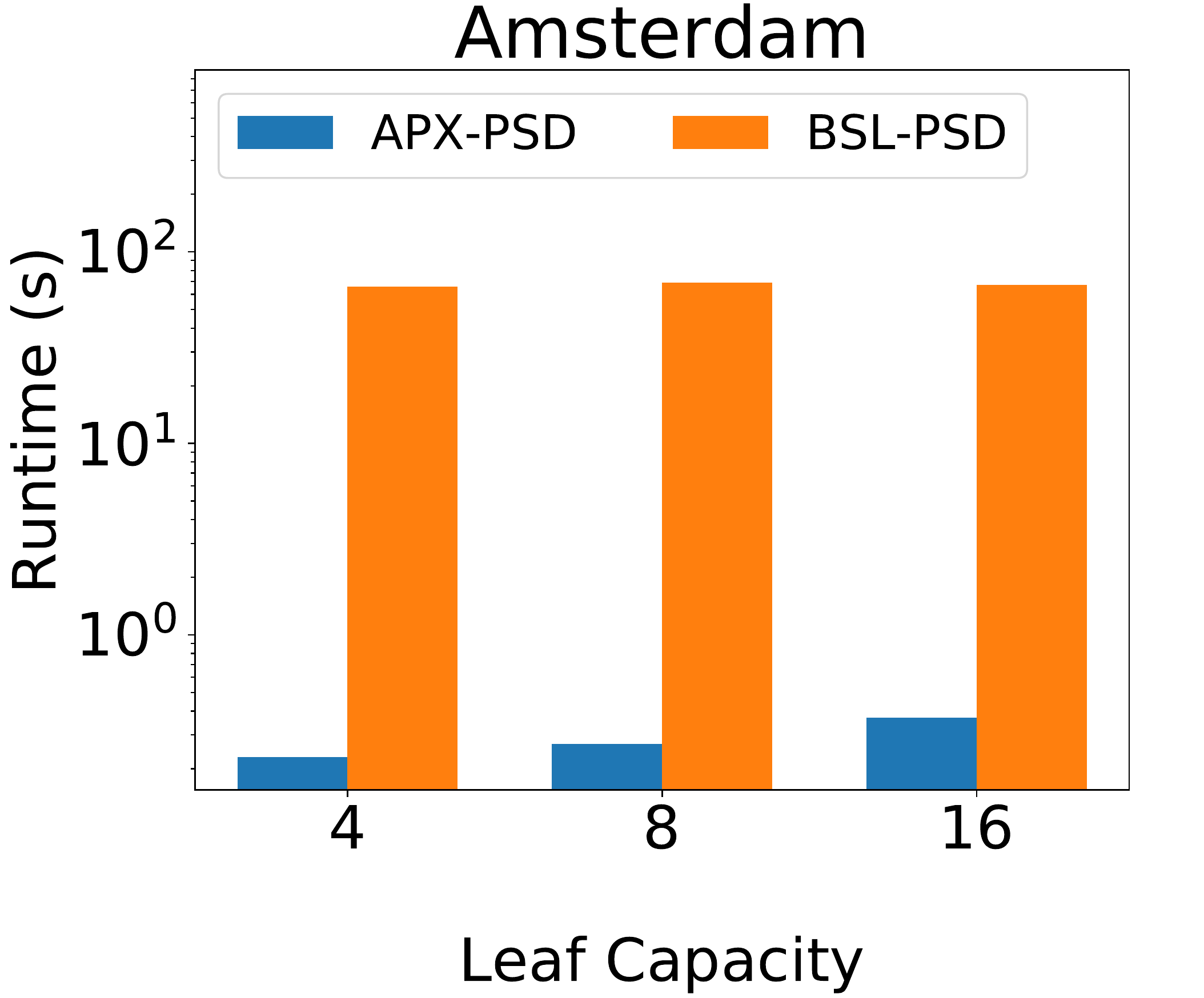}
        \includegraphics[width=0.49\columnwidth]{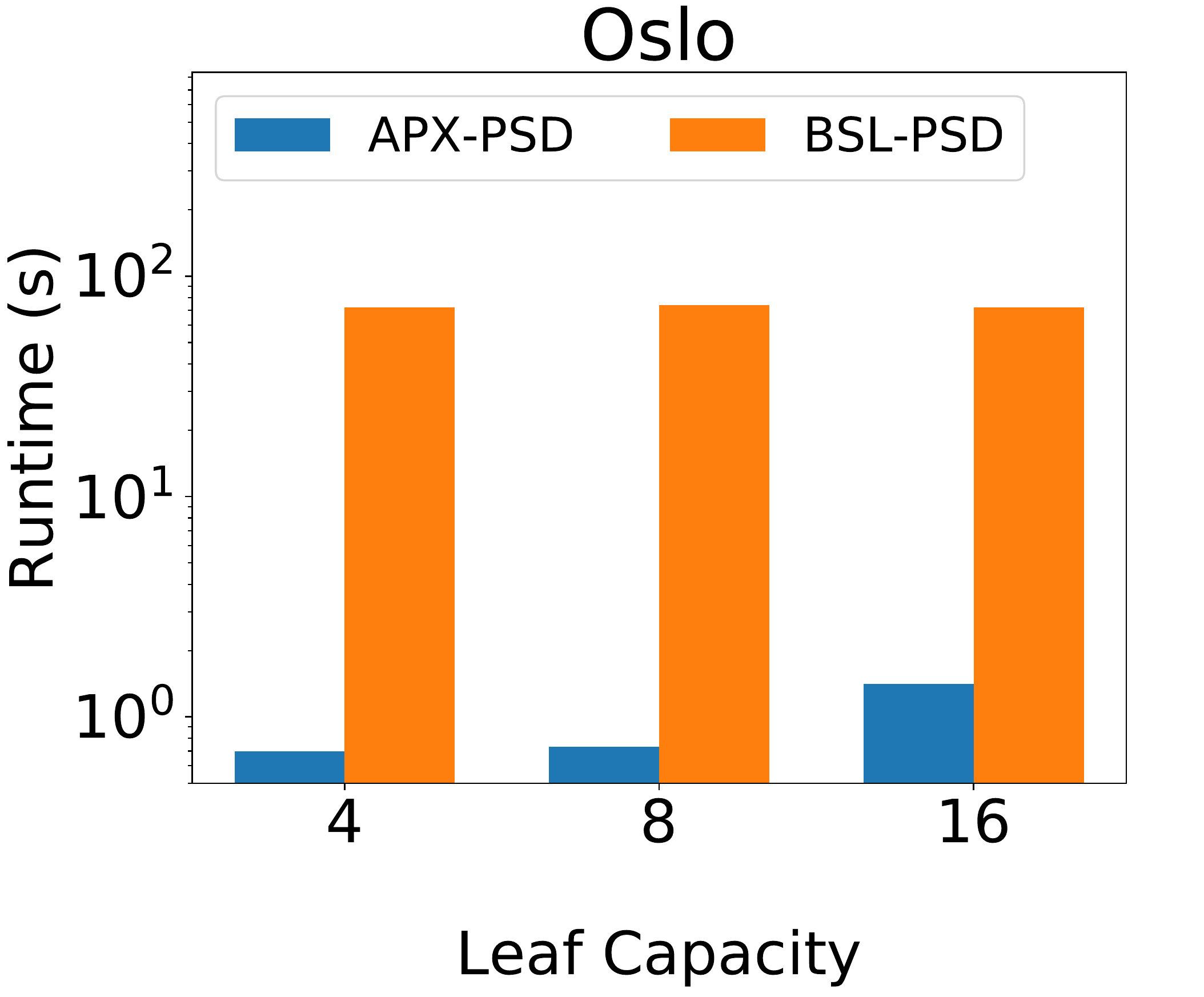}
        \includegraphics[width=0.49\columnwidth]{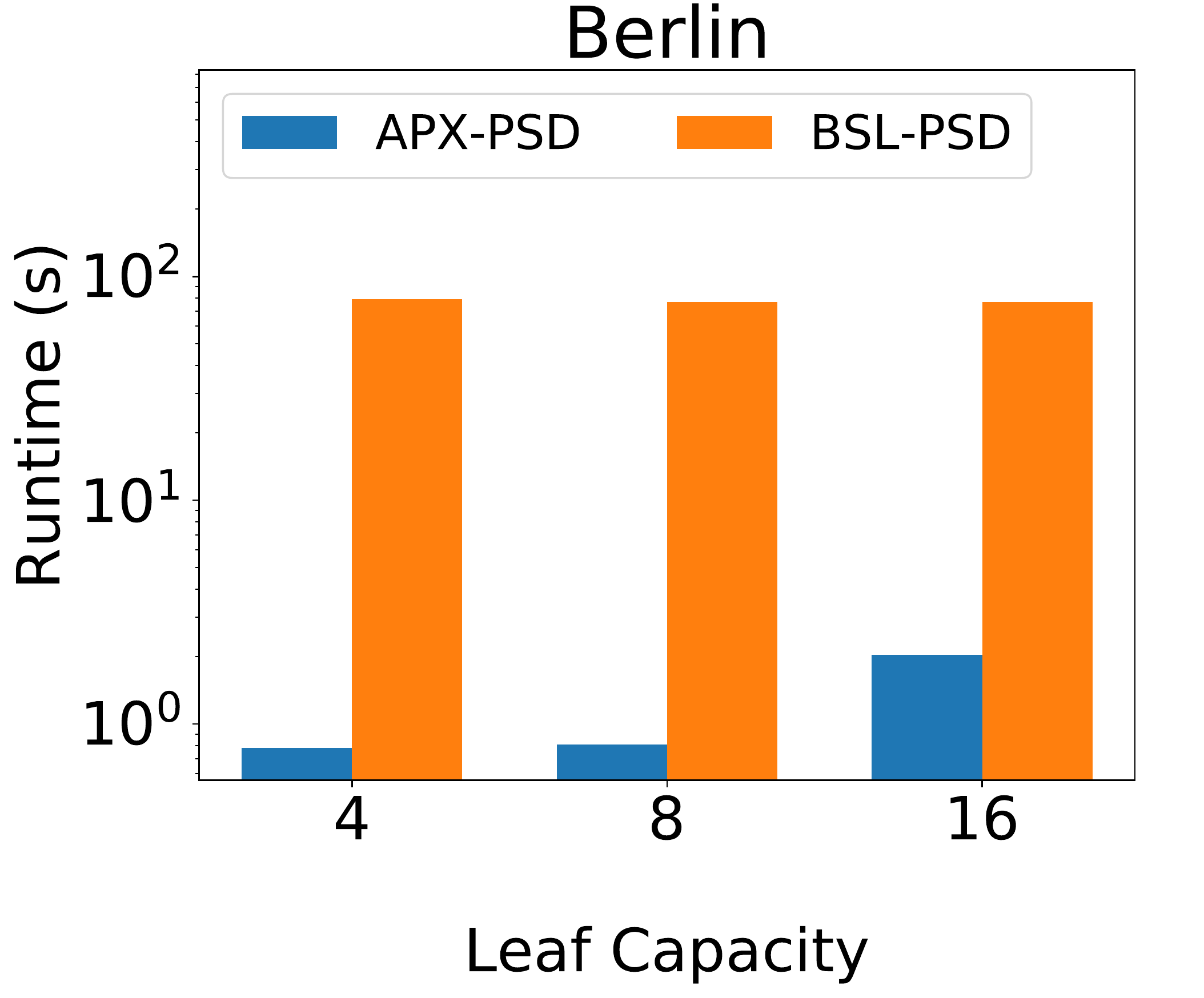}
   \caption{{Efficiency w.r.t. with varying quad-tree leaf size }}
   \label{fig:t_leaf}
\end{figure}

Finally, Figure \ref{fig:t_leaf} shows that APX-\NameProblem's efficiency decreases as the leaf capacity increases, due to the increased number of partial shopping routes that APX-\NameProblem generates and evaluates from the leaves it visits -- indeed, from APX-\NameProblem's complexity analysis (Sec. \ref{appendix: appx complexity}) recall that this represents the time-dominant component of this approach.

%% file: 2_related.tex
\section{Related work}
\label{section:relatedwork}

To the extent of our knowledge there is no prior work in the literature that address the \NameProblem problem as presented here.  Nonetheless, in the following we discuss some works which are somehow related.

One can think of \NameProblem of an extension of the well-known Optimal Sequence Routing (OSR) \cite{OSR} and Trip Planning Queries (TPQs) \cite{Li2005}.
An OSR query \cite{OSR} specifies a sequence of categories-of-interest (COIs) and returns the optimal route that visits exactly one point-of-interest (POI) from each COI in the exact sequence provided in the query. 
Trip Plannng Queries (TPQs) \cite{Li2005} are similar to OSR queries where the strict sequence of COIs is relaxed.
%
Unfortunately, solutions to TPQs and OSR queries are not applicable to \NameProblem queries, particularly because \NameProblem queries consider two competing cost criteria, whereas in both TPQs and ORS queries only travel cost (distance) is considered.
%

The problem of finding the set of routes that are optimal under a given combination of cost criteria has also been researched the skyline paradigm, e.g., \cite{Kriegel2010}), however none of those works have a notion similar to fulfilling a shopping list while building a route, as in the \NameProblem query, but merely aim at finding routes.

In the context of our work the literature targeting \textit{bicriteria} networks \cite{Raith2009} is particularly relevant.  
%
%
Linear skylines were employed in various \textit{instances} of bicriteria networks \cite{Shekelyan2014, Ahmadi2017, Costa2018}, each requiring to devise a specific solution addressing the peculiarities of the associated problem setting, which, again, are all different from \NameProblem's query.

%% file: 6_conclusions.tex
\section{Conclusion}

\label{sec: conclusions}

In this paper we proposed a solution to what we called the ``Personal Shopper's Dilemma'' query, which is to decide on how to fulfil a customer's shopping list while minimizing travel time as well as shopping cost.  The idea is to leave the shopper to decide, ``on the fly'' how to prioritize these two criteria.  Given the query's NP-hardness we proposed a heuristic solution that leverages on the concept of linear skyline queries.  In order to measure the effectiveness of the proposed heuristic we also proposed a metric to evaluate its loss w.r.t. an optimal solution.  Using real city-scale datasets we show that our proposal is able to deliver good linear skylines yielding optimality and coverage gaps below 50\% and 15\% respectively
two orders of magnitude faster than the optimal solution.  

A direction for future work w.r.t. the \NameProblem query itself would be allowing a shopper to find routes to serve multiple customers, possibly in different locations, and/or have multiple shoppers that could, for instance, bid on shopping lists of different customers after considering their perspective on those two criteria.
